\documentclass[12pt]{article}
\usepackage{graphicx}
\usepackage{amsmath,amssymb,amsfonts,amsthm}

\usepackage{setspace}
\usepackage{caption}
\usepackage{booktabs}
\usepackage{hyperref}
\usepackage{color}
\usepackage[stable]{footmisc}
\hypersetup{
	pdftitle={What Impulse Response Do Instrumental Variables Identify?},
	pdfauthor={Bonsoo Koo, Seojeong Lee, Myung Hwan Seo},
	colorlinks,%
	citecolor=black,%
	filecolor=black,%
	linkcolor=black,%
	urlcolor=blue
}
\makeatletter
\let\FN@baselinestretch\baselinestretch
\AtBeginDocument{{\footnotesize\global\footnotesep=0.7\baselineskip}}
\makeatother
\makeatletter
\renewcommand\@biblabel[1]{}
\renewcommand\@openbib@code{\setlength{\leftmargin}{1.5em}\setlength{\itemindent}{-1.5em}\setlength{\labelwidth}{0pt}\setlength{\labelsep}{0pt}}
\makeatother

\usepackage[top=1in, bottom=1in, left=1in, right=1in]{geometry}

\usepackage{titlesec}
\titleformat{\section}{\Large\bfseries}{\thesection}{1em}{}

\newtheorem{remark}{Remark}
\newtheorem{proposition}{Proposition}
\newtheorem{corollary}{Corollary}
\newtheorem{assumption}{Assumption}

\DeclareMathOperator*{\cov}{cov}
\DeclareMathOperator*{\var}{var}

\theoremstyle{definition}
\newtheorem{assump}{Assumption}
\newenvironment{myassump}[2][]
{\renewcommand\theassump{#2}\begin{assump}[#1]}
	{\end{assump}}

\usepackage{newtxtext,newtxmath}

\begin{document}
	
	\begin{titlepage}
\newgeometry{top=1in,bottom=1in,left=1in,right=1in}

\title{\Large{What Impulse Response Do Instrumental Variables Identify?}\thanks{We acknowledge that this research is supported under the Australian Research Council Discovery Project funding scheme (project number DP210101440) and by the Seoul National University Research Grant in 2021.}}
\author{Bonsoo Koo\thanks{Department of Econometrics and Business Statistics, Monash University, bonsoo.koo@monash.edu} \and Seojeong Lee\thanks{Department of Economics, Seoul National University, s.jay.lee@snu.ac.kr} \and Myung Hwan Seo\thanks{Department of Economics, Hong Kong University of Science and Technology, myunghseo@ust.hk}}
\date{First version: August 25, 2022\thanks{arXiv:2208.11828}\\ This version: October 2026}
\maketitle
\begin{abstract}
We study what local projection instrumental variable (LP-IV) regressions identify when an instrument is
correlated with several structural shocks, such as defense and non-defense spending shocks. LP-IV identifies an
instrument-specific affine combination of the component impulse responses, possibly with negative weights.
Over-identified 2SLS or GMM yields another such combination, and the over-identification
test can reject even with valid instruments. With as many instruments as shocks, the component responses are
identified when each shock has a policy variable (e.g., sectoral spending) and its impact on the other policy
variables is restricted. We illustrate with government spending and tax multipliers.
\end{abstract}
\begin{quotation}\small\noindent Keywords: local projection, instrumental variables, structural vector moving average, impulse response,
negative weights, government spending multiplier, tax multiplier.
\end{quotation}
\end{titlepage}
\addtocounter{page}{1}
\newgeometry{top=1in,bottom=1in,right=1in,left=1in}

\section{Introduction}\label{Intro}

Empirical macroeconomists routinely use instrumental variables within local projections (LP-IV) to estimate impulse responses to policy shocks. The standard interpretation is clearest when the instrument isolates one structural shock. In practice, however, what the instrument isolates is often a combination of several structural shocks. A government spending shock is by construction the sum of defense and non-defense spending shocks, and a monetary policy surprise combines the pure policy shock with a central bank information shock, which is not the object of interest. When the instrument is correlated with more than one of these underlying shocks, what does the LP-IV estimand identify? We show that it is an affine combination of the impulse responses to the underlying shocks, with weights that can be negative, so that even an instrument that is valid for the policy shock as a whole need not identify any single structural response, nor a convex combination of them.

We formalize this in a structural vector moving average model in which the instrument can be correlated with several of the underlying structural shocks. The LP-IV estimand is then a weighted sum of the responses to the individual structural shocks that make up the policy shock, which we call the component impulse responses. Each weight is proportional to the covariance between the instrument and that shock. These weights are unrestricted in sign. Existing LP-IV analyses characterize this estimand only when the instrument is correlated with a single structural shock (Stock and Watson, 2018; Plagborg-M\o ller and Wolf, 2021).

As an example, consider the cumulative government spending multiplier of Ramey and Zubairy (2018), estimated by LP-IV with narrative military news as the instrument. Since government spending is the sum of defense and non-defense spending, the LP-IV estimand $\beta_{h}$, which measures the cumulative response of real GDP up to horizon $h$ per unit of the cumulative response of government spending, decomposes as
\begin{equation}
	\label{decom0}
	\beta_{h} = w_{\text{defense}}\times \theta_{h,\text{defense}} + w_{\text{non-defense}}\times\theta_{h,\text{non-defense}},
\end{equation}
where $ \theta_{h,\text{defense}}$ and $\theta_{h,\text{non-defense}}$ are the defense and non-defense spending multipliers, respectively, and the weights $w_{\text{defense}}$ and $w_{\text{non-defense}}$ have signs that depend on the correlation between the military news instrument and the corresponding sectoral spending shock. The estimand $\beta_{h}$ is a proper weighted average of the two sectoral multipliers only when the weights are between zero and one, in which case it can be interpreted as an aggregate spending multiplier. However, when we apply the specification of Ramey and Zubairy (2018) to the post-WWII sample in Section~\ref{Sec-Gov}, we find that for $h=18$ (eighteen quarters), under the assumption that each sector's shock does not affect the other sector's spending and with the weights estimated from sectoral spending as in Section~\ref{Sec-GD}, the estimated decomposition is
	\begin{equation}
	\begin{array}{ccccccccc}
		\widehat{\beta}_{h} &= & \widehat{w}_{\text{defense}}&  \times& \widehat{\theta}_{h,\text{defense}} &+&  \widehat{w}_{\text{non-defense}}&\times&\widehat{\theta}_{h,\text{non-defense}}.\\
		0.37& &1.87 &&0.68 & & -0.87 & &1.02 \\
	\end{array}
\end{equation}
Due to the negative weight, the estimated government spending multiplier is smaller than both the defense and the non-defense spending multipliers. This can mislead researchers and policymakers even when the instrument and the estimation procedure are valid. Ramey and Zubairy themselves recognize that the defense and non-defense responses may differ and relate the issue to the distinction between the local average treatment effect (LATE) and the average treatment effect (ATE). Their remark, however, is not a formal analysis of the identification problem. This paper provides that analysis.

The issue arising from the negative weights has received much attention recently in the microeconometrics literature. For instance, de Chaisemartin and D’Haultfoeuille (2020) show that the regression coefficient in the two-way fixed effects model can be expressed as a weighted sum of the average treatment effects with weights that can be negative because treatment timing differs across units. Mogstad, Torgovitsky, and Walters (2021) show that the two-stage least squares (2SLS) estimand using multiple instruments may lose a causal interpretation due to negative weights assigned to some subpopulations under general treatment effect heterogeneity. Goldsmith-Pinkham, Hull, and Koles\'{a}r (2024) show that in a regression on multiple treatments, each coefficient is contaminated by the other treatments' effects with contamination weights that can be negative.

We develop the local-projection counterpart, in which the components are structural shocks and the object of interest is a dynamic impulse response. In the time-series setting, Koles\'{a}r and Plagborg-M\o ller (2025) show that local projections on an observed shock or a proxy estimate a weighted average of marginal effects when the response to a single structural shock is nonlinear. In our setting, by contrast, the negative weights arise from a single instrument that is correlated with several structural shocks, and they appear even when every impulse response is linear.

Using several instruments together in an over-identified regression for the aggregate regressor (for example, total spending), as in 2SLS or generalized method of moments (GMM), which we call over-identified LP-GMM, does not identify the component responses. Ramey and Zubairy (2018) report 2SLS estimates using both the narrative military news and spending shock instruments, but the 2SLS estimand is an affine combination of the single-instrument estimands with weights that depend on the correlation between the instruments and can be negative. The resulting LP-2SLS and LP-GMM estimands can therefore lack a structural interpretation, and the over-identifying restrictions test cannot distinguish invalid instruments from a composite policy shock.

A negative-weight estimand loses its interpretation as the causal effect of the aggregate shock but still carries information about the underlying responses. When several instruments are available, each yields a different affine combination of the same component responses, so the component responses could be solved from these combinations if the weights were known.

The weights, however, depend on the covariances between each instrument and the unobserved component shocks, so they are not identified from the single-instrument estimands alone. They can be identified from the policy variable of each component, such as defense and non-defense spending. We call this variable the component's normalizing variable because the shock moves it one-for-one on impact. The covariance between an instrument and each normalizing variable combines the instrument's correlations with the component shocks through the effect of each shock on the other normalizing variable. Once these inter-component impact responses are fixed, the weights and the component responses are identified. In sum, identification requires (i) a normalizing variable for each component, (ii) as many instruments as components, whose correlations with the component shocks differ, and (iii) a restriction on the inter-component impact responses. Under these conditions the signs of the weights no longer matter, and the decomposition applies to existing single-instrument LP-IV estimates using only the covariances of the instruments with the normalizing variables.

The restriction on the inter-component impact responses is the identifying assumption, and it appears in the literature under other names. The recursive ordering of a proxy SVAR and exclusion restrictions on the instruments each imply particular values of the inter-component impact responses. Sections~\ref{Sec-GD-ord} and \ref{Sec-GD-excl} derive these values for the recursive ordering of Mertens and Ravn (2013) and for exclusion restrictions, including the projection-residual instruments of Miranda-Agrippino and Ricco (2021).

The normalizing variables need not sum to an aggregate. In our second application, the average personal and corporate income tax rates of Mertens and Ravn (2013) serve as the normalizing variables for the two tax shocks, with their narrative tax measures as the instruments, and we compare the responses implied by their recursive ordering with those implied by exclusion restrictions.

Beyond the work already discussed, our paper relates to a large literature that uses IVs in macroeconometrics. Stock and Watson (2012), Mertens and Ravn (2013), Gertler and Karadi (2015), and Jaroci\'{n}ski and Karadi (2020) employ IVs for identification in SVAR models. Giacomini, Kitagawa, and Read (2022) and Arias, Rubio-Ram\'{i}rez, and Waggoner (2021) study proxy SVARs with several proxies for several shocks, in which the proxies are combined with additional restrictions. The LP method was introduced by Jord\`{a} (2005), and LP-IV has been used in several studies, such as Jord\`{a}, Schularick, and Taylor (2015, 2020), Stock and Watson (2018), and Ramey and Zubairy (2018). For a recent overview of LP-IV, we refer to Jord\`{a} and Taylor (2025). Rambachan and Shephard (2025) provide a nonparametric causal interpretation of these estimands. Miranda-Agrippino and Ricco (2021) propose an instrument construction that reduces the bias when an instrument is correlated with several shocks.

Our findings have implications beyond LP-IV. Since we characterize the population estimand, the result applies to any estimator of the same object. Plagborg-M\o ller and Wolf (2021) show that local projections and VARs estimate the same impulse responses, including with an instrument, so the same cautions apply to impulse responses obtained from VARs with the instrument ordered first. 

The rest of the paper proceeds as follows. Section~\ref{Sec-ID} presents the model and the characterization of the LP-IV estimand, Section~\ref{Sec-2SLS} shows that over-identified LP-GMM, which uses several instruments for the aggregate regressor, does not identify the component responses, Section~\ref{Sec-GD} develops identification with normalizing variables, Section~\ref{Sec-Gov} applies the results to the U.S. government spending multiplier of Ramey and Zubairy (2018), Section~\ref{Sec-MR} revisits the personal and corporate income tax multipliers of Mertens and Ravn (2013), and Section~\ref{Sec-Conclusion} concludes. Appendix~\ref{Sec-pf} contains the proofs and Appendix~\ref{Sec-LATE} the comparison with the potential outcomes framework.

\section{Model and the LP-IV Estimand}
\label{Sec-ID}

To motivate our modeling choice, consider a researcher who estimates the government spending multiplier in a bivariate SVAR:
\begin{equation*}
A\left(L\right)\left(\begin{array}{c}
	x_{t}\\
	y_{t}
\end{array}\right)=\left(\begin{array}{c}
	\varepsilon_{xt}\\
	\varepsilon_{yt}
\end{array}\right),	
\end{equation*}
where $x_{t}$ is government spending, $y_{t}$ is GDP, $(\varepsilon_{xt},\varepsilon_{yt})$ are mutually uncorrelated structural shocks, and $A(L)$ is the lag polynomial. The government spending shock is the sum of a defense and a non-defense spending shock, $\varepsilon_{xt}=\varepsilon_{xt}^{D}+\varepsilon_{xt}^{ND}$, and the two components plausibly generate different responses of GDP. The SVAR cannot accommodate this heterogeneity. Within the system, the spending shock $\varepsilon_{xt}$ has a single impulse response by construction, and letting the two components enter as separate shocks with their own responses yields three structural shocks in a two-variable system. The shocks can then no longer be recovered from current and past observables, so the system is not invertible. We therefore adopt a structural vector moving average (SVMA) model, which allows more structural shocks than observed variables and does not require invertibility.

\subsection{Structural Vector Moving Average}

Let $\boldsymbol{Y}_{t}$ be an $n\times1$ vector of observed endogenous variables and let $\boldsymbol{\varepsilon}_{t}$ be an $m\times1$ vector of unobserved structural shocks. The vector of endogenous variables is a linear combination of current and past shocks:
\begin{equation}
	\boldsymbol{Y}_{t} = \boldsymbol{\Theta}(L)\boldsymbol{\varepsilon}_{t}
	\label{svma}
\end{equation}
where $L$ is the lag operator, $\boldsymbol{\Theta}(L) = \boldsymbol{\Theta}_{0} + \boldsymbol{\Theta}_{1}L + \boldsymbol{\Theta}_{2}L^{2}+\cdots$, and $\boldsymbol{\Theta}_{h}$ for $h=0,1,2,\ldots$ is an $n\times m$ matrix of impulse response coefficients. We assume that the $\boldsymbol{\varepsilon}_{t}$'s are i.i.d. over time, $E[\boldsymbol{\varepsilon}_{t}]=0$, and $E[\boldsymbol{\varepsilon}_{t}\boldsymbol{\varepsilon}_{t}']\equiv \boldsymbol{\Lambda}>0$ where $\boldsymbol{\Lambda}$ is a diagonal matrix (i.e., the shocks are mutually uncorrelated). We reserve the index $s$ for the elements of $\boldsymbol{\varepsilon}_{t}$, so that $\varepsilon_{s,t}$ denotes the $s$-th structural shock. 

Let $y_{t}$ be the last element of $\boldsymbol{Y}_{t}$ without loss of generality. We define the impulse response of $y_{t}$ at horizon $h$ to the shock $\varepsilon_{s,t}$ as
\begin{equation*}
	\theta_{h,ys}\equiv E[y_{t+h}|\varepsilon_{s,t}=1] - E[y_{t+h}|\varepsilon_{s,t}=0].
\end{equation*}
Here we are slightly abusing notation by letting $\theta_{h,ys}$ be the $(n,s)$-th element of $\boldsymbol{\Theta}_{h}$. This is to emphasize that $y_{t}$ is the main variable of interest. 

 Suppose that a researcher is interested in the impulse response of $y_{t}$ at horizon $h$ to a macroeconomic policy shock associated with several structural shocks in $\boldsymbol{\varepsilon}_{t}$. Without loss of generality, the policy shock comprises the first $S$ structural shocks, $\varepsilon_{1,t},\ldots,\varepsilon_{S,t}$, which we call its components.
Proposition~\ref{P1} below characterizes what the LP-IV estimand identifies when the instrument is correlated with several component shocks. In the LP-IV literature, such a policy shock is often treated as if it were generated by a single structural shock, which implicitly rules out heterogeneous component responses.

A number of papers document the composite nature of macroeconomic policy shocks, and the composition matters in two distinct ways. For fiscal policy, the component shocks are well defined by the sectoral decomposition of spending, and the parameters of interest are the responses to the component shocks or a convex combination of them, which is an aggregate multiplier. Auerbach and Gorodnichenko (2012) show that the responses to disaggregated spending shocks differ from those to the aggregate fiscal policy shock, and Cox, M\"uller, Pastén, Schoenle, and Weber (2024) and Bouakez, Rachedi, and Santoro (2025) show that the composition of government spending materially affects the aggregate multiplier. For monetary policy, the decomposition instead reflects the distinct nature of the shocks contained in a policy announcement, and the components need not be of equal interest. In Jaroci\'{n}ski and Karadi (2020), the shock of interest is the pure monetary policy shock. The announcement surprise also contains a central bank information shock, and the empirical goal is to remove it. In Kaminska, Mumtaz, and \v{S}ustek (2021), the announcement comprises action, expected path, and uncertainty shocks, and the responses to each component are of interest. In both cases, applied work often targets a policy shock whose underlying shock composition is not observed.

The policy shock under study is typically unobserved and its scale is indeterminate. A common solution is to measure shock magnitude by means of an observable endogenous variable $x_{t}$ and then to use an IV $z_{t}$ correlated with $x_{t}$ to obtain exogenous variation. As a result, the response of $y_{t+h}$ to $z_{t}$ relative to the response of $x_{t}$ to $z_{t}$ is interpreted as the average response of $y_{t+h}$ to a policy shock scaled to move $x_{t}$ by one unit. Stock and Watson (2018) provide an example in which the response of GDP growth ($y_{t}$) to a monetary policy shock is identified by the ratio of the impulse responses of GDP growth and the federal funds rate ($x_{t}$) to a monetary policy announcement ($z_{t}$). Ramey and Zubairy (2018) calculate the government spending multiplier as a ratio of the impulse responses of GDP ($y_{t}$) and total government spending ($x_{t}$) to the military news shock ($z_{t}$).

We take $x_{t}$ to be the first element of $\boldsymbol{Y}_{t}$, again without loss of generality. Similar to $y_{t}$, we define the impulse response of $x_{t}$ at horizon $h$ to the shock $\varepsilon_{s,t}$ as $\theta_{h,xs}$, slightly abusing notation. We assume $\theta_{0,xs}>0$ for all $s = 1, 2, \dots, S$, which we refer to as the sign normalization, throughout the paper. Where indicated, we also fix the scale of each shock by the unit effect normalization, $\theta_{0,xs}=1$ for all $s=1,2,\dots,S$ (Stock and Watson, 2018), so that responses are measured per unit impact effect on $x_{t}$. In the government spending example, the unit effect normalization implies that a one-unit change in a sectoral spending shock results in a one-unit change in total spending, so the component shocks are also expressed in common units. Section~\ref{Sec-GD} instead scales each component shock by its nonzero impact on its own normalizing variable.

\subsection{Local Projections with an Instrumental Variable}

Let $z_{t}$ be an IV that satisfies the following assumptions.
\begin{assumption}\
	\begin{enumerate}
		\item[(i)] $\cov(x_{t},z_{t})\neq0$ (relevance)
		\item[(ii)] $\cov(z_{t},\varepsilon_{s,t})=0$ for all $s=S+1, S+2,\ldots,m$ (contemporaneous exogeneity)
		\item[(iii)] $\cov(z_{t},\boldsymbol{\varepsilon}_{t+j})=\boldsymbol{0}$ for $j\neq0$ (lead-lag exogeneity)
	\end{enumerate}
	\label{A-IV}
\end{assumption}
\noindent
Define $\alpha_{s} = \cov(z_{t},\varepsilon_{s,t})$. Under Assumptions~\ref{A-IV}(ii) and (iii), $\cov(x_{t},z_{t})=\sum_{s=1}^{S}\alpha_{s}\theta_{0,xs}$, so Assumption~\ref{A-IV}(i) implies that $\alpha_{s}\neq0$ for some $s=1,2,\ldots,S$. Assumption~\ref{A-IV} is an extension of Condition LP-IV of Stock and Watson (2018) allowing more than one structural shock to be correlated with the IV, with heterogeneous correlations across the shocks.

The following proposition establishes that the LP-IV estimand is an affine combination of the normalized component impulse responses $\theta_{h,ys}/\theta_{0,xs}$, which are the responses per unit impact on $x_{t}$. The proof is in Appendix~\ref{Sec-pf}.
\begin{proposition}
	If random variables $y_{t}$ and $x_{t}$ are elements of $\boldsymbol{Y}_{t}$ generated according to \eqref{svma} and a random variable $z_{t}$ satisfies Assumption~\ref{A-IV}, and $\theta_{0,xs}>0$ for $s=1,2,\ldots,S$, then for $h=0,1,2,\ldots$,\begin{equation}
		\label{P1-e1}
		\beta_{h}\equiv \frac{\cov(y_{t+h},z_{t})}{\cov(x_{t},z_{t})} = \sum_{s=1}^{S}w_{s}\frac{\theta_{h,ys}}{\theta_{0,xs}},
	\end{equation}where
	\begin{equation*}
		w_{s} = \frac{\alpha_{s}\theta_{0,xs}}{\sum_{s'=1}^{S}\alpha_{s'}\theta_{0,xs'}}.
	\end{equation*}
	\label{P1}
	\vspace{-1em}
\end{proposition}
Proposition~\ref{P1} implies that the effect identified by LP-IV is instrument-specific when the instrument $z_t$ is correlated with at least two component shocks ($S\geq2$) with heterogeneous impulse responses. 

When the instrument is correlated with only one structural shock (i.e., $S=1$), \eqref{P1-e1} simplifies to Equation (8) of Stock and Watson (2018). In this case, the LP-IV estimand $\beta_{h} =\theta_{h,y1}$ under the unit effect normalization $\theta_{0,x1}=1$. Thus, Proposition~\ref{P1} generalizes the previous identification result to cases in which the instrument is correlated with multiple shocks in the SVMA model \eqref{svma}.\footnote{Technically, our model can equivalently handle situations where an IV is correlated with an arbitrary set of shocks in the SVMA model, which would typically render it an invalid IV in conventional settings.}

\begin{remark}
\textup{
The SVMA model offers a structural interpretation of the impulse responses $\theta_{h,ys}$ in Proposition~\ref{P1}. However, a result similar to \eqref{P1-e1} can be obtained even without relying on a structural model.\footnote{We thank an anonymous referee for this insightful comment.} Define $\theta_{h,ys}$ as the coefficient from the orthogonal projection, given by $\cov(y_{t+h},\varepsilon_{s,t})/\var(\varepsilon_{s,t})$. Suppose $z_{t}$ is a linear combination of the component shocks: $z_t = \lambda'\boldsymbol{e}_{t}$, where $\boldsymbol{e}_{t} \equiv (\varepsilon_{1,t}, \varepsilon_{2,t}, \cdots, \varepsilon_{S,t})'$. Since the shocks $\varepsilon_{s,t}$ are mutually uncorrelated, the orthogonal projection of $y_{t+h}$ on $\boldsymbol{e}_{t}$ is $\mathcal{P}(y_{t+h}|\boldsymbol{e}_{t}) = \sum_{s=1}^{S} \theta_{h,ys} \varepsilon_{s,t}$. The projection error $y_{t+h} - \mathcal{P}(y_{t+h}|\boldsymbol{e}_{t})$ is orthogonal to the subspace spanned by $\boldsymbol{e}_{t}$, and hence also to $z_t$. Therefore, the projection of $y_{t+h}$ onto $z_t$ is:
\begin{align*}
\mathcal{P}(y_{t+h}|z_{t}) &= \mathcal{P}\left(\mathcal{P}(y_{t+h}|\boldsymbol{e}_{t})|z_{t}\right) + \mathcal{P}\left(y_{t+h} - \mathcal{P}(y_{t+h}|\boldsymbol{e}_{t})|z_{t}\right) \\
&= \underbrace{\left(\sum_{s=1}^{S} \theta_{h,ys} \frac{\cov(z_{t}, \varepsilon_{s,t})}{\var(z_{t})} \right)}_{\text{projection coefficient}} z_{t}.
\end{align*}
Multiplying both sides by $\var(z_{t})$ yields $\cov(y_{t+h}, z_t) = \sum_{s=1}^{S} \theta_{h,ys} \cov(z_{t}, \varepsilon_{s,t})$, which corresponds to the numerator in \eqref{P1-e1}. For example, a classical measurement error in the instrument can be treated as a component $\varepsilon_{S,t}$ with $\theta_{h,yS}=0$. This calculation shows that even in general projection models, the LP-IV estimand can be expressed as a linear combination of underlying projection coefficients, highlighting the broader applicability of Proposition~\ref{P1}.
}
\end{remark}

\smallskip

Proposition~\ref{P1} characterizes the estimand, but it is not an identification result for the component responses. The right-hand side of \eqref{P1-e1} is not a convex combination, which has nonnegative weights, but an affine combination whose weights sum to one and can be negative. Let $S=2$. Then \eqref{P1-e1} becomes
\begin{equation}
\label{ID0}
\beta_{h} =\frac{\alpha_{1}\theta_{0,x1}}{\alpha_{1}\theta_{0,x1}+\alpha_{2}\theta_{0,x2}} \times \frac{\theta_{h,y1}}{\theta_{0,x1}} + \frac{\alpha_{2}\theta_{0,x2}}{\alpha_{1}\theta_{0,x1}+\alpha_{2}\theta_{0,x2}} \times \frac{\theta_{h,y2}}{\theta_{0,x2}}.
\end{equation}
It is clear from \eqref{ID0} that without restrictions on the signs of $\alpha_{s}$, we cannot interpret $\beta_{h}$ as a structural parameter, as it could take any real value. For example, suppose $\theta_{h,y1}=1$, $\theta_{h,y2}=2$, and $\theta_{0,x1} = \theta_{0,x2} = 1$, so the normalized impulse responses are both positive. However, if $\alpha_{1} = 0.2$ and $\alpha_{2} = -0.1$, then $\beta_{h} = 0$. Such a situation arises when the IV is correlated with the two structural shocks, $\varepsilon_{1,t}$ and $\varepsilon_{2,t}$, with opposite signs.

Under the following assumption, we may interpret $\beta_{h}$ as a meaningful average of the underlying component impulse responses. 
\begin{myassump}{\textit{Same-Sign}}
	\textit{Either $\alpha_{s}\theta_{0,xs}\geq0$ for all $s=1,2,\ldots,S$, or $\alpha_{s}\theta_{0,xs}\leq0$ for all $s=1,2,\ldots,S$.}
	\label{A-mono}
\end{myassump}
\noindent
This assumption imposes a requirement beyond Assumption~\ref{A-IV} for the LP-IV estimand to have a structural (causal) interpretation. Given the sign normalization for $\theta_{0,xs}$'s, this is equivalent to imposing the same sign conditions on the $\alpha_s$'s. Economic theory can often justify this assumption for a given IV and its relation to the component shocks within the policy shock. For example, Jarociński and Karadi (2020) use high-frequency surprises in fed funds futures ($z_{t}^{ff}$) and in stock prices ($z_{t}^{sp}$) as the IVs for monetary policy announcements that contain both a pure monetary policy shock ($\varepsilon_{1,t}$) and a central bank information shock ($\varepsilon_{2,t}$) in the SVAR model. Under the sign normalization, they impose a restriction that $\alpha_{1}>0$ and $\alpha_{2}>0$ for $z_{t}^{ff}$, but $\alpha_{1}<0$ and $\alpha_{2}>0$ for $z_{t}^{sp}$. Since $z_{t}^{ff}$ satisfies Assumption~\ref{A-mono}, the impulse response identified by $z_{t}^{ff}$ is a convex combination of the responses to the two component shocks, while the estimand based on $z_{t}^{sp}$ identifies neither the pure monetary policy response nor such a combination.

\begin{remark}
	\textup{
	In the empirical work of Stock and Watson (2012), which uses a dynamic factor model, seventeen external instruments are considered to identify six structural shocks: oil shock, monetary policy shock, productivity shock, uncertainty shock, liquidity and financial risk shock, and fiscal policy shock. Among them, the Romer and Romer (2010) instrument is intended to identify a tax component of a fiscal policy shock, whereas the Ramey (2011) and the Fisher and Peters (2010) instruments are intended to identify spending components. They estimate the fiscal policy shock using each IV separately and find a correlation of $-0.93$ between the shocks estimated with the Fisher and Peters and the Romer and Romer instruments, although the two instruments are intended for different components. This suggests that each instrument is correlated with more than one component of the fiscal policy shock.
}
\end{remark}

\begin{remark}
	\textup{
	There are cases in which an LP-IV estimand with negative weights may itself be the parameter of interest. McKay and Wolf (2023) show that, for counterfactual policy-rule analysis, the causal effects of contemporaneous and news shocks to the prevailing policy rule can be sufficient inputs. When all component shocks are shocks to the policy rule, $\beta_{h}$ is the response to a particular policy package and, under their assumptions, can be used for such analysis, whatever the signs of the weights. This case does not apply to our examples, where the parameters of interest are the responses to each component shock separately.
}
\end{remark}

Stock and Watson (2018) interpret the impulse response as the dynamic causal effect of a shock, with the shock in the role of the treatment. This connects the LP-IV estimand to program evaluation, where, under treatment effect heterogeneity, a valid instrument identifies a local average treatment effect (Imbens and Angrist, 1994). The LATE framework does not transfer directly to our setting. The treatment is the continuous composition of the shock rather than a binary or finite treatment, and the response is homogeneous given that composition, so the heterogeneity is in the treatment rather than in the effect. Assumption \textit{Same-Sign} plays the role of the Imbens-Angrist monotonicity condition in keeping the weights nonnegative. Monotonicity restricts individual potential treatment choices, whereas Assumption \textit{Same-Sign} restricts the covariances between the instrument and the component shocks. Rambachan and Shephard (2025) give a nonparametric causal interpretation of the LP-IV estimand when the instrument is correlated with a single shock. We characterize the case in which it is correlated with several. A detailed comparison with the potential outcomes framework is in Appendix~\ref{Sec-LATE}.

The SVMA model \eqref{svma} applies to mean-zero variables, typically differences or growth rates, while researchers may be interested in the level of $\boldsymbol{Y}_{t}$. The difference in the levels between $t$ and $t+h$ can be written as the cumulative changes over the $h$ periods: $\sum_{j=0}^{h}\boldsymbol{Y}_{t+j}$. Write $\widetilde{y}_{t+h} = \sum_{j=0}^{h}y_{t+j}$ and $\widetilde{x}_{t+h} = \sum_{j=0}^{h}x_{t+j}$. Define the cumulative impulse responses as $\widetilde{\theta}_{h,ys}=\sum_{j=0}^{h}\theta_{j,ys}$ and $\widetilde{\theta}_{h,xs}=\sum_{j=0}^{h}\theta_{j,xs}$, respectively. 

\begin{corollary}
	\label{C1}
	If random variables $y_{t}$ and $x_{t}$ are elements of $\boldsymbol{Y}_{t}$ generated according to \eqref{svma}, and a random variable $z_{t}$ satisfies Assumptions~\ref{A-IV}(ii) and (iii), then for $h=0,1,2,\ldots$ such that $\cov(\widetilde{x}_{t+h},z_{t})\neq0$ and $\widetilde{\theta}_{h,xs}\neq0$ for $s=1,2,\ldots,S$,
	\begin{equation}
		\label{C1-e1}
		\widetilde{\beta}_{h}\equiv\frac{\cov\left(\widetilde{y}_{t+h},z_{t}\right)}{\cov\left(\widetilde{x}_{t+h},z_{t}\right)} = \sum_{s=1}^{S}\left(\frac{\alpha_{s}\widetilde{\theta}_{h,xs}}{\sum_{s'=1}^{S}\alpha_{s'}\widetilde{\theta}_{h,xs'}}\right)\frac{\widetilde{\theta}_{h,ys}}{\widetilde{\theta}_{h,xs}}.
	\end{equation}
\end{corollary}

Corollary~\ref{C1} is relevant for identification and estimation of the cumulative impulse responses, such as the cumulative government spending multiplier of Ramey and Zubairy (2018). Their LP-IV estimand takes the form of the left-hand side of \eqref{C1-e1} after controlling for the lagged variables, where $y_{t}$ is GDP, $x_{t}$ is government spending, and $z_{t}$ is the military news shock, each divided by trend GDP. 

For the cumulative estimand in Corollary~\ref{C1}, the relevant version of Assumption \textit{Same-Sign} applies to the products $\alpha_{s}\widetilde{\theta}_{h,xs}$, whose signs may vary with $h$. In the government spending example, with defense ($s=1$) and non-defense ($s=2$) spending shocks, all signs but one are readily justified. A positive sectoral spending shock raises cumulative total spending, $\widetilde{\theta}_{h,xs}>0$ for $s=1,2$ and $h\geq0$, and the military news shock is positively correlated with the defense spending shock, $\alpha_{1}>0$. Whether the estimand is a convex combination of the cumulative sectoral multipliers then depends on the sign of $\alpha_2$. We find empirical evidence supporting $\alpha_{2}<0$ in Section~\ref{Sec-Gov}, under the assumption that each sector's shock does not affect the other sector's spending.

\section{Over-identified LP-GMM Does Not Identify the Component Responses}\label{Sec-2SLS}

In the previous section, we characterized the LP-IV estimand when a single IV is correlated with several component shocks. In practice, several IVs for the same policy shock are often available, and it is common practice to use all of them together for efficiency. The regression of $y_{t+h}$ on the single regressor $x_{t}$ is then over-identified, and its 2SLS and GMM estimators use all the IVs for $x_{t}$. We call these estimators over-identified LP-GMM, with LP-2SLS as the leading case. Stock and Watson (2018) consider LP-GMM with several instruments, and Ramey and Zubairy (2018) estimate the government spending multiplier by LP-2SLS with two IVs.\footnote{Ramey and Zubairy (2018) call these the estimates with ``\textit{combined instruments}''. Their Stata code shows that they are 2SLS estimates.}

Proposition~\ref{P1} implies that the over-identified regression does not identify the component responses. The 2SLS estimand is a further affine combination of the single-IV estimands, and the combination weights can themselves be negative. When the instruments' correlations with the component shocks differ, so that the single-IV estimands differ, the LP-2SLS estimand need not have a structural interpretation even when every IV is valid and each single-IV estimand is a convex combination of the component responses.

We make this point precise using the result of Andrews (2019). Consider two instruments, $z_{t}^{A}$ and $z_{t}^{B}$, and suppose the policy shock under study has two component shocks, $\varepsilon_{1,t}$ and $\varepsilon_{2,t}$. By Proposition~\ref{P1}, the LP-IV estimand is expressed as
\begin{equation}
	\beta_{h}^{j} = w_{1}^{j}\frac{\theta_{h,y1}}{\theta_{0,x1}} + (1-w_{1}^{j})\frac{\theta_{h,y2}}{\theta_{0,x2}},
	\label{eq-2sec}
\end{equation}
for $j=A,B$, where $\beta^{A}_{h}$ and $\beta^{B}_{h}$ are the LP-IV estimands using $z_{t}^{A}$ and $z_{t}^{B}$ one at a time, respectively, and $w_{1}^{j}$ is the weight of Proposition~\ref{P1} with $\alpha_{s}^{j}=\cov(z_{t}^{j},\varepsilon_{s,t})$. Let $\beta^{2SLS}_{h}$ be the 2SLS estimand using both instruments together. According to Andrews (2019), $\beta^{2SLS}_{h} = p\beta^{A}_{h} + (1-p)\beta^{B}_{h}$, where the scalar $p$ is a known function of the second moments of the instruments and $x_{t}$ and is not restricted to $[0,1]$.\footnote{Let $Z_{t} = (z_{t}^{A},z_{t}^{B})'$. Then $p = E[Z_{t}x_{t}]'\left(E[Z_{t}Z_{t}']\right)^{-1}\left(\begin{array}{cc} E[z_{t}^{A}x_{t}] & 0 \end{array}\right)' \big/ \left(E[Z_{t}x_{t}]'\left(E[Z_{t}Z_{t}']\right)^{-1}E[Z_{t}x_{t}]\right)$.} Substituting this result into \eqref{eq-2sec}, we get:
\begin{equation}
	\label{eq-2sls}
	\beta^{2SLS}_{h} = \left\{pw_{1}^{A} + (1-p)w_{1}^{B}\right\}\frac{\theta_{h,y1}}{\theta_{0,x1}} + \left\{p(1-w_{1}^{A}) + (1-p)(1-w_{1}^{B})\right\}\frac{\theta_{h,y2}}{\theta_{0,x2}}.
\end{equation}
Here, the weights sum to one but can be negative. Since $p$ depends on the correlation between the IVs, $\beta^{2SLS}_{h}$ may not be a proper weighted average even when $z_{t}^{A}$ and $z_{t}^{B}$ satisfy Assumption~\ref{A-mono}, in which case each LP-IV estimand $\beta_{h}^{A}$ and $\beta_{h}^{B}$ is causal. The example below illustrates such a case. 

\bigskip

\noindent\textbf{Example: a non-causal 2SLS estimand as an affine combination of causal LP-IV estimands}\newline
Suppose that for $-1<\rho<1$,
\begin{align*}
	E[Z_{t}Z_{t}'] =& \left(\begin{array}{cc}
		1 & \rho \\
		\rho & 1 \\
	\end{array}\right), ~~E[Z_{t}x_{t}] = \left(\begin{array}{c}
		E[z_{t}^{A}x_{t}] \\
		E[z_{t}^{B}x_{t}] 
	\end{array}\right) = \left(\begin{array}{c}
		1 \\
		1.625 \\
	\end{array}\right)\\
	\alpha_{1}^{A} = & 0.5, ~~ \alpha_{2}^{A} =0.5,~~\alpha_{1}^{B} =  0.125, ~~ \alpha_{2}^{B} =  1.5.
\end{align*}
Assumption~\ref{A-mono} is satisfied for both IVs. We calculate that $w_{1}^{A} = 1/2$, $w_{1}^{B}=1/13$, and
\[\beta^{2SLS}_{h} = \frac{15(3-4\rho)\theta_{0,x1}}{233-208\rho}\cdot\frac{\theta_{h,y1}}{\theta_{0,x1}} +  \frac{4(47-37\rho)\theta_{0,x2}}{233-208\rho}\cdot\frac{\theta_{h,y2}}{\theta_{0,x2}}.\]
Then the weight for $\theta_{h,y1}/\theta_{0,x1}$ is negative if $3/4<\rho<1$.

\bigskip

Another implication of Proposition~\ref{P1} concerning the use of 2SLS with multiple IVs is that the over-identifying restrictions test (also known as Hansen's J test or the Sargan test) rejects the null hypothesis of jointly valid IVs in large samples when the single-IV estimands differ. The test asks whether the IV estimates based on a single IV converge in probability to a unique value, but this is not the case if each IV identifies a different weighted average (Lee, 2018). To be precise, the null hypothesis is given by the moment condition $H_{0}: 0 = E[Z_{t}(y_{t+h}-x_{t}\beta^{*})]$ where $\beta^{*}$ is the presumed true value estimated by 2SLS. Thus, the null hypothesis of the test can be written as 
\begin{equation}
H_{0}: ~0 = E[z_{t}^{A}(y_{t+h}-x_{t}\beta_{h}^{2SLS})]= E[z_{t}^{B}(y_{t+h}-x_{t}\beta_{h}^{2SLS})].
\end{equation}
However, $\beta_{h}^{A}$ and $\beta_{h}^{B}$ are the unique values that satisfy $0=E[z_{t}^{A}(y_{t+h}-x_{t}\beta_{h}^{A})]$ and $0=E[z_{t}^{B}(y_{t+h}-x_{t}\beta_{h}^{B})]$.\footnote{Since $E[y_{t+h}]=0$ in the SVMA model \eqref{svma}, $\cov(y_{t+h},z_{t})=E[y_{t+h}z_{t}]$. The same applies to $x_{t}$.} Therefore, the null hypothesis of the test does not hold when $\beta_{h}^{A}\neq\beta_{h}^{B}$. Indeed, Ramey and Zubairy (2018) report p-values of the test between 0.03 and 0.05 in their footnote 22. The rejection may be due to either the invalidity of some of the IVs under the conventional assumption (e.g., Stock and Watson, 2018) or the composite nature of the policy shock. To the best of our knowledge, there is no test to distinguish the two in the macroeconometrics literature.
\section{Identification of Component Responses}\label{Sec-GD}

Rather than adding instruments for the single regressor, this section adds one observable variable per component shock. The single-IV estimands contain the information needed to identify the component responses, but their weights depend on the unobserved correlations $\alpha_{s}^{j}$ between the instruments and the component shocks. The added variables reveal these correlations once the inter-component impact responses are fixed. We call each of them the normalizing variable of its shock. It is the policy variable that the shock moves one-for-one on impact, as $x_{t}$ scales the policy shock in Section~\ref{Sec-ID}. In the government spending application, the normalizing variables are defense and non-defense spending, which are the disaggregated series of $x_{t}$. In the tax application, they are the personal and corporate income tax rates. We state identification in two equivalent forms, as a regression on the normalizing variables and as a system in the single-IV estimands of Section~\ref{Sec-ID}.

Formally, the augmented SVMA model includes the additional normalizing variables in the system:
\begin{equation}
	\boldsymbol{Y}_{t}^{A} = \boldsymbol{\Psi}(L)\boldsymbol{\varepsilon}_{t},
	\label{svma2}
\end{equation}
where $\boldsymbol{Y}_{t}^{A} = (x_{1,t},x_{2,t},\ldots,x_{S,t},\ldots,y_{t})'$ replaces $x_{t}$ in $\boldsymbol{Y}_{t}$ with $S$ observed normalizing variables, one for each component shock, $\boldsymbol{\Psi}(L)=\boldsymbol{\Psi}_{0} + \boldsymbol{\Psi}_{1}L + \boldsymbol{\Psi}_{2}L^{2}+\cdots$, and $\boldsymbol{\Psi}_{h}$ is the $(n+S-1)\times m$ matrix of impulse responses. We write $\psi_{h,rs}$ for the $(r,s)$-th element of $\boldsymbol{\Psi}_{h}$, which is the response of $x_{r,t+h}$ to the component shock $\varepsilon_{s,t}$, and normalize $\psi_{0,ss}=1$ for $s=1,2,\ldots,S$, so that the unit effect normalization now fixes the scale of each component shock by its own normalizing variable. 
When the normalizing variables are the disaggregated series of $x_{t}$, aggregating the first $S$ rows of \eqref{svma2} by $x_{t}=\sum_{s=1}^{S}x_{s,t}$ recovers the baseline model \eqref{svma}, with $\boldsymbol{\Theta}(L)=\boldsymbol{B}\boldsymbol{\Psi}(L)$ for the $n\times(n+S-1)$ aggregation matrix
\[
\boldsymbol{B}=
	\begin{pmatrix}
		\boldsymbol{1}_{S}' & \boldsymbol{0}_{1\times(n-1)}\\
		\boldsymbol{0}_{(n-1)\times S} & I_{n-1}
	\end{pmatrix},
\] 
where 	$\boldsymbol{1}_{S}$ is the $S\times1$ vector of ones and $\boldsymbol{0}_{i\times j}$ is the $i\times j$ matrix of zeros.
Then, $\theta_{h,xs}=\sum_{r=1}^{S}\psi_{h,rs}$. The rows for the variables other than $x_{t}$, including $y_{t}$, are unchanged, so the responses $\theta_{h,ys}$ are the same objects as in \eqref{svma}. Section~\ref{Sec-ID} measures responses per unit impact on $x_{t}$, whereas here each shock is scaled by its own normalizing variable, so the two scales differ by the factor $\theta_{0,xs}=\sum_{r}\psi_{0,rs}$. Without further restrictions, \eqref{svma2} is more general than \eqref{svma}.

From here on we set $S=2$, which covers our applications. The extension to general $S$ is discussed after Proposition~\ref{P-aug}.

A key identifying restriction is on the inter-component impact responses, $\psi_{0,rs}$ for $r\neq s$ in \eqref{svma2}. These are the impact responses of $x_{r,t}$ to a unit increase in $\varepsilon_{s,t}$, and they have precise economic interpretations. For example, let $\varepsilon_{1,t}$ and $\varepsilon_{2,t}$ be the defense and the non-defense spending shock, respectively. If $\psi_{0,21}=-0.2$, then a unit increase in the defense spending shock increases defense spending by the same amount ($\psi_{0,11}=1$) but decreases non-defense spending by 0.2. When $\psi_{0,12}=\psi_{0,21}=0$, the defense shock does not affect non-defense spending on impact, and vice versa.

To see how this restriction achieves identification, observe that under Assumption~\ref{A-IV}, $\cov(x_{s,t},z_{t}) = \alpha_{s}+\psi_{0,sr}\alpha_{r}$ for $s\neq r$. By solving for $\alpha_{s}$, we obtain
	\begin{equation*}
		\begin{pmatrix} \alpha_{1} \\ \alpha_{2} \end{pmatrix}
		=\Psi_{0}^{-1}\begin{pmatrix} \cov(x_{1,t},z_{t}) \\ \cov(x_{2,t},z_{t}) \end{pmatrix},
	\end{equation*}
where $\Psi_{0}=\bigl(\begin{smallmatrix}1 & \psi_{0,12}\\ \psi_{0,21} & 1\end{smallmatrix}\bigr)$ denotes the upper-left $2\times2$ block of $\boldsymbol{\Psi}_{0}$, which contains the impact responses of $(x_{1,t},x_{2,t})$ to $(\varepsilon_{1,t},\varepsilon_{2,t})$ and is assumed to be invertible. Given $\psi_{0,12}$ and $\psi_{0,21}$, the coefficients $\alpha_{s}$ are identified. Since $\Psi_{0}$ does not depend on the IV, the same restriction identifies the coefficients $\alpha_{s}$ of a second IV. Since $\cov(y_{t+h},z_{t})=\alpha_{1}\theta_{h,y1}+\alpha_{2}\theta_{h,y2}$ for each IV, combining the two IVs then identifies $\theta_{h,y1}$ and $\theta_{h,y2}$.

Define 
\begin{equation*}
	C=\left[\cov(x_{s,t},z_{t}^{j})\right]_{j,s},\qquad
	\boldsymbol{c}_{h}=\left[\cov(y_{t+h},z_{t}^{j})\right]_{j},
\end{equation*}
where the rows correspond to the instruments $j=A,B$ and the columns of $C$ to the components $s=1,2$, and let $\boldsymbol{b}_{h}=C^{-1}\boldsymbol{c}_{h}$, which is the 2SLS estimand of the local projection of $y_{t+h}$ on $(x_{1,t},x_{2,t})$ with $(z_{t}^{A},z_{t}^{B})$ as the IVs. Unlike the over-identified regression of Section~\ref{Sec-2SLS}, which uses multiple instruments for one endogenous variable, this regression is just identified. In addition, for any random variables $a_{t}^{A}$ and $a_{t}^{B}$ with $\cov(a_{t}^{j},z_{t}^{j})\neq0$, define the single-IV LP-IV estimands and weights
\begin{equation*}
	\beta_{h}^{j}=\frac{\cov(y_{t+h},z_{t}^{j})}{\cov(a_{t}^{j},z_{t}^{j})},\qquad
	w_{s}^{j}=\frac{\cov(x_{s,t},z_{t}^{j})}{\cov(a_{t}^{j},z_{t}^{j})},\qquad
	W=\left[w_{s}^{j}\right]_{j,s}.
\end{equation*}
    The coefficients $w_{s}^{j}$ coincide with the weights of Proposition~\ref{P1} when $x_{t}=x_{1,t}+x_{2,t}$, $a_{t}^{j}=x_{t}$, and $\psi_{0,12}=\psi_{0,21}=0$. The following proposition formally establishes identification of $\theta_{h,y1}$ and $\theta_{h,y2}$ using the normalizing variables $x_{1,t}$ and $x_{2,t}$ and a restriction on $\psi_{0,sr}$, $s\neq r$.

\begin{proposition}\label{P-aug}
	Suppose that random variables $y_{t}$, $x_{1,t}$, and $x_{2,t}$ are elements of $\boldsymbol{Y}_{t}^{A}$ generated according to the augmented SVMA model \eqref{svma2} with $S=2$, that $\psi_{0,21}$ and $\psi_{0,12}$ are given with $\psi_{0,12}\psi_{0,21}\neq1$, that two IVs $z_{t}^{A}$ and $z_{t}^{B}$ satisfy Assumptions~\ref{A-IV}(ii) and (iii), and that $C$ is invertible. Then, for any choice of $a_{t}^{A}$ and $a_{t}^{B}$,
	\begin{equation*}
		\left(\begin{array}{c} \theta_{h,y1}\\ \theta_{h,y2}\end{array}\right)
		= \Psi_0'\boldsymbol{b}_{h}
		= \Psi_0'W^{-1}\left(\begin{array}{c} \beta_{h}^{A}\\ \beta_{h}^{B}\end{array}\right).
	\end{equation*}
\end{proposition}
	\bigskip

The two equalities give the component responses in two forms. The first expresses them as the coefficients of the just-identified 2SLS regression on the normalizing variables, adjusted by the inter-component impact responses. The tax application uses this form. The second decomposes the single-IV estimates that applied work reports, and its weights $W$ are the objects needed to check Assumption \textit{Same-Sign}. The government spending application uses the second form with the two instruments, each with its own control variables as in Ramey and Zubairy (2018). Either form accommodates instrument-specific controls by residualizing each instrument on its own controls, provided that each instrument satisfies Assumptions~\ref{A-IV}(ii) and (iii) after this residualization and that the controls are predetermined, so that the component responses and $\Psi_{0}$ are the same across the regressions.

Invertibility of $C$ is the standard rank condition for identification in models with two endogenous variables and two IVs. In our case, $C$ is not invertible if the correlations of the two IVs with the component shocks are proportional, for example if both IVs are correlated with the same component shock only. The choice of the regressors $a_{t}^{j}$ does not matter because $\cov(a_{t}^{j},z_{t}^{j})$ divides both $\beta_{h}^{j}$ and the weights $w_{s}^{j}$, and cancels in $W^{-1}(\beta_{h}^{A},\beta_{h}^{B})'$. The first equality of the proposition does not involve $a_{t}^{j}$ at all. When $\psi_{0,12}=\psi_{0,21}=0$, $(\theta_{h,y1},\theta_{h,y2})'=\boldsymbol{b}_{h}$, so the just-identified 2SLS regression on the normalizing variables estimates the component responses directly. 

We use two choices of $a_{t}^{j}$. The first is $a_{t}^{A}=a_{t}^{B}=x_{t}$ when the normalizing variables are the disaggregated series of $x_{t}$. Then each $\beta_{h}^{j}$ is the LP-IV estimand with the aggregate regressor, and each row of $W$ sums to one because the covariances of the normalizing variables with $z_{t}^{j}$ add up to $\cov(x_{t},z_{t}^{j})$. The second is $a_{t}^{A}=x_{1,t}$ and $a_{t}^{B}=x_{2,t}$, which pairs each instrument with its own normalizing variable and sets the diagonal of $W$ to one.

	For general $S$, the same argument requires an observable series for each component and $S$ valid IVs. Proposition~\ref{P-aug} then applies as stated, with the $S\times S$ matrices $C$ and $\Psi_{0}$ in place of the $2\times2$ matrices. The rank condition is again the invertibility of $C$, which requires the instruments' correlations with the component shocks to be linearly independent.

	The government spending application uses cumulative changes over horizons $0$ to $h$. Recall from the end of Section~\ref{Sec-ID} that $\widetilde{y}_{t+h}=\sum_{\ell=0}^{h}y_{t+\ell}$, $\widetilde{x}_{t+h}=\sum_{\ell=0}^{h}x_{t+\ell}$, $\widetilde{\theta}_{h,ys}=\sum_{\ell=0}^{h}\theta_{\ell,ys}$, and $\widetilde{\theta}_{h,xs}=\sum_{\ell=0}^{h}\theta_{\ell,xs}$, and let $\widetilde{x}_{r,t+h}=\sum_{\ell=0}^{h}x_{r,t+\ell}$ and $\widetilde{\psi}_{h,rs}=\sum_{\ell=0}^{h}\psi_{\ell,rs}$, the response of $\widetilde{x}_{r,t+h}$ to $\varepsilon_{s,t}$, so that $\widetilde{\theta}_{h,x1}=\widetilde{\psi}_{h,11}+\widetilde{\psi}_{h,21}$ and $\widetilde{\theta}_{h,x2}=\widetilde{\psi}_{h,22}+\widetilde{\psi}_{h,12}$. The parameter of interest is the cumulative component response $\widetilde{\theta}_{h,ys}/\widetilde{\psi}_{h,ss}$, which is the response of $\widetilde{y}_{t+h}$ to $\varepsilon_{s,t}$ per unit cumulative change in $x_{s,t}$. An example is the cumulative non-defense spending multiplier. The following proposition is the cumulative version of Proposition~\ref{P-aug} with $a_{t}^{A}=a_{t}^{B}=\widetilde{x}_{t+h}$.

	\begin{proposition} \label{P-aug-cumul}
	Suppose that random variables $y_{t}$, $x_{1,t}$, and $x_{2,t}$ are elements of $\boldsymbol{Y}_{t}^{A}$ generated according to the augmented SVMA model \eqref{svma2} with $S=2$ and $x_{t}=x_{1,t}+x_{2,t}$. For a given horizon $h$, suppose that $\widetilde{\psi}_{h,11}\neq0$, $\widetilde{\psi}_{h,22}\neq0$, the relative cumulative inter-component impulse responses $\widetilde{\psi}_{h,21}/\widetilde{\psi}_{h,11}$ and $\widetilde{\psi}_{h,12}/\widetilde{\psi}_{h,22}$ are given and $(\widetilde{\psi}_{h,21}/\widetilde{\psi}_{h,11})(\widetilde{\psi}_{h,12}/\widetilde{\psi}_{h,22})\neq1$. In addition, suppose that two IVs $z_{t}^{A}$ and $z_{t}^{B}$ satisfy Assumptions~\ref{A-IV}(ii) and (iii), $\cov(\widetilde{x}_{t+h},z_{t}^{j})\neq0$ for $j=A,B$, and $\widetilde{W}=[\widetilde{w}_{s}^{j}]_{j,s}$ is invertible, where $\widetilde{w}_{s}^{j}=\cov(\widetilde{x}_{s,t+h},z_{t}^{j})/\cov(\widetilde{x}_{t+h},z_{t}^{j})$. Let $\widetilde{\beta}_{h}^{j}=\cov(\widetilde{y}_{t+h},z_{t}^{j})/\cov(\widetilde{x}_{t+h},z_{t}^{j})$, and let $\widetilde{\boldsymbol{b}}_{h}$ be $\boldsymbol{b}_{h}$ with $\widetilde{y}_{t+h}$ and $\widetilde{x}_{s,t+h}$ in place of $y_{t+h}$ and $x_{s,t}$. Then $(\widetilde{\theta}_{h,y1}/\widetilde{\psi}_{h,11},\widetilde{\theta}_{h,y2}/\widetilde{\psi}_{h,22})$ is identified as
		\begin{equation*}
		\left(\begin{array}{c}
			\widetilde{\theta}_{h,y1}/\widetilde{\psi}_{h,11} \\ 
			\widetilde{\theta}_{h,y2}/\widetilde{\psi}_{h,22}
		\end{array}\right)
		=   \left(\begin{array}{cc}
			1 & \widetilde{\psi}_{h,21}/\widetilde{\psi}_{h,11} \\
			\widetilde{\psi}_{h,12}/\widetilde{\psi}_{h,22} & 1 \\
		\end{array}\right)\widetilde{\boldsymbol{b}}_{h},\qquad
		\widetilde{\boldsymbol{b}}_{h}=\widetilde{W}^{-1}	\left(
		\begin{array}{c}
			\widetilde{\beta}_{h}^{A} \\ 
			\widetilde{\beta}_{h}^{B} \\ 
		\end{array}\right).
	\end{equation*}
	\end{proposition}

The proof is analogous to that of Proposition~\ref{P-aug}, with $\widetilde{\Psi}_{h}=[\widetilde{\psi}_{h,rs}]$ in place of $\Psi_{0}$ and the cumulative covariances in place of $C$ and $\boldsymbol{c}_{h}$. Dividing $\widetilde{\boldsymbol{\theta}}_{h}=(\widetilde{\theta}_{h,y1},\widetilde{\theta}_{h,y2})'=\widetilde{\Psi}_{h}'\widetilde{\boldsymbol{b}}_{h}=\widetilde{\Psi}_{h}'\widetilde{W}^{-1}(\widetilde{\beta}_{h}^{A},\widetilde{\beta}_{h}^{B})'$ by $\widetilde{\psi}_{h,ss}$ row by row gives the result, and the invariance to the choice of $a_{t}^{j}$ applies as well.

In sum, the decomposition of LP-IV estimands into the component responses requires three things: (i) a normalizing variable for each component shock, (ii) as many valid and relevant IVs as components, whose correlations with the component shocks differ enough for $C$ to be invertible, and (iii) a restriction on the inter-component impact responses or, for cumulative responses, on their cumulative counterparts. 

Fixing the values of the inter-component impact responses is an identifying restriction on $\Psi_{0}$, and our framework connects this restriction explicitly to identification strategies used in practice. Any strategy that point-identifies the component shocks implies particular values of the inter-component impact responses, so the results under each strategy are the results at the value it implies. The following subsections derive the implied values for the recursive ordering in the proxy SVAR of Mertens and Ravn (2013) and for exclusion restrictions on the IVs, including the projection residual of Miranda-Agrippino and Ricco (2021).

\subsection{Recursive Ordering in the Proxy SVAR}\label{Sec-GD-ord}

Let $\boldsymbol{\nu}_{t}$ be the innovation of $\boldsymbol{Y}_{t}^{A}$, defined as the error of the linear projection of $\boldsymbol{Y}_{t}^{A}$ on its own past, and partition it as $\boldsymbol{\nu}_{t}=(\boldsymbol{\nu}_{t}^{c\,\prime},\boldsymbol{\nu}_{t}^{o\,\prime})'$, where $\boldsymbol{\nu}_{t}^{c}$ corresponds to the normalizing variables and $\boldsymbol{\nu}_{t}^{o}$ to the remaining variables. Let $\boldsymbol{\varepsilon}_{t}^{c}=(\varepsilon_{1,t},\varepsilon_{2,t})'$ be the component shocks, let $\Lambda_{c}=\var(\boldsymbol{\varepsilon}_{t}^{c})$ denote the block of $\boldsymbol{\Lambda}$ for the component shocks, and define $\Gamma=E[\boldsymbol{\nu}_{t}\boldsymbol{\varepsilon}_{t}^{c\,\prime}]$ with blocks $\Gamma^{c}$ and $\Gamma^{o}$ conformable with the partition of $\boldsymbol{\nu}_{t}$. The projection on the past is a linear function of lagged $\boldsymbol{Y}_{t}^{A}$, hence of past shocks by \eqref{svma2}, so
\begin{equation*}
	\nu_{r,t}=\sum_{s}\psi_{0,rs}\,\varepsilon_{s,t} + \{\boldsymbol{\varepsilon}_{t-1},\boldsymbol{\varepsilon}_{t-2},\cdots\},
\end{equation*}
where $\{\cdots\}$ denotes a linear combination of past shocks. Since the shocks are mutually uncorrelated, $\Gamma^{c}=\Psi_{0}\Lambda_{c}$, whether or not the system is invertible.

The proxy SVAR adds the assumption that the component shocks are invertible in the sense of Miranda-Agrippino and Ricco (2023). Each component shock is then a linear combination of the current innovations. As in Mertens and Ravn (2013, equation (15)), we write
\begin{equation}\label{svarcorr1}
	\boldsymbol{\nu}_{t}^{c}=\eta\,\boldsymbol{\nu}_{t}^{o}+H\boldsymbol{\varepsilon}_{t}^{c}
\end{equation}
for some matrices $\eta$ and nonsingular $H$. Nonsingularity of $H$ is slightly stronger than invertibility of the component shocks, since it also requires that no linear combination of them is a function of $\boldsymbol{\nu}_{t}^{o}$ alone. The remaining shocks may be noninvertible, unlike in the fully invertible SVAR of Mertens and Ravn (2013). Taking the covariance of \eqref{svarcorr1} with $\boldsymbol{\varepsilon}_{t}^{c}$ gives $\Gamma^{c}=\eta\Gamma^{o}+H\Lambda_{c}$. Substituting $\Gamma^{o}=\zeta\Gamma^{c}$ with $\zeta=\Gamma^{o}(\Gamma^{c})^{-1}$ and $\Gamma^{c}=\Psi_{0}\Lambda_{c}$, and canceling $\Lambda_{c}$,
\begin{equation}\label{svarcorr2}
\Psi_{0}=F\,H,
\end{equation}
where $F=\left(I_{2}-\eta\zeta\right)^{-1}$. Mertens and Ravn (2013) show that the covariances between the IVs and the innovations, together with the covariance matrix of the innovations, identify $\zeta$ and $\eta$, hence $F$ and $\Omega_{c}=\var(\boldsymbol{\nu}_{t}^{c}-\eta\boldsymbol{\nu}_{t}^{o})=H\Lambda_{c}H'$, but not $H$ itself. The remaining unknowns are the four elements of $H$ and the two variances in $\Lambda_{c}$. They satisfy five equations. Three come from $\Omega_{c}=H\Lambda_{c}H'$ and two from the normalization $\psi_{0,ss}=1$, which sets the diagonal of $FH$ to one. With one more restriction on $H$, $\Psi_{0}$ is fixed.

The recursive ordering of Mertens and Ravn (2013) with $x_{1,t}$ first takes $H$ to be the Cholesky factor of $\Omega_{c}$ when the shocks are normalized to unit variance. Since $\Lambda_{c}$ is diagonal, the same restriction in our normalization is $H_{12}=0$ in \eqref{svarcorr1}. Then \eqref{svarcorr2} and $\psi_{0,ss}=1$ give $\psi_{0,22}=F_{22}H_{22}=1$, $\psi_{0,12}=F_{12}H_{22}$, $\psi_{0,11}=F_{11}H_{11}+F_{12}H_{21}=1$, and $\psi_{0,21}=F_{21}H_{11}+F_{22}H_{21}$. Hence, for $F_{22}\neq0$ and $F_{11}+F_{12}k\neq0$,
\begin{equation}\label{svarcorr3}
	\psi_{0,12}=\frac{F_{12}}{F_{22}},\qquad
	\psi_{0,21}=\frac{F_{21}+F_{22}\,k}{F_{11}+F_{12}\,k},\qquad
	k=\frac{H_{21}}{H_{11}}=\frac{[\Omega_{c}]_{21}}{[\Omega_{c}]_{11}},
\end{equation}
where the last equality uses $\Omega_{c}=H\Lambda_{c}H'$ with $H_{12}=0$. The reversed ordering sets $H_{21}=0$ and implies another value by the same steps.

\subsection{Exclusion Restrictions on the IVs}\label{Sec-GD-excl}

An exclusion restriction assumes that an IV is correlated with only one of the two component shocks. The restriction can be imposed on the IV itself or on a transformation of it. Miranda-Agrippino and Ricco (2021) construct their instrument for monetary policy shocks as the residual of a projection of market surprises on Greenbook forecasts and on past surprises. The projection removes the information effect, so that the residual measures the monetary policy shock alone. In the same way, we consider the residual of the projection of one IV on the other. Each restriction implies a particular value of one inter-component impact response.

Let $z_{t}^{A}$ be the IV for $\varepsilon_{1,t}$. Let $\tilde{z}_{t}^{A}=z_{t}^{A}-\delta z_{t}^{B}$ be the residual of its projection on $z_{t}^{B}$, where $\delta=\cov(z_{t}^{A},z_{t}^{B})/\var(z_{t}^{B})$.
\begin{myassump}{\textit{Exclusion}}
	\textit{$z_{t}^{A}$ satisfies Assumption~\ref{A-IV} with $S=1$ and $x_{1,t}$ in place of $x_{t}$.}
	\label{A-excl}
\end{myassump}
\begin{myassump}{\textit{Residual-Exclusion}}
	\textit{$\tilde{z}_{t}^{A}$ satisfies Assumption~\ref{A-IV} with $S=1$ and $x_{1,t}$ in place of $x_{t}$.}
	\label{A-rexcl}
\end{myassump}
\noindent
Assumption~\ref{A-excl} is the usual requirement that an IV is correlated with a single shock. Assumption~\ref{A-rexcl} applies when the researcher is concerned that $z_{t}^{A}$ is correlated with $\varepsilon_{2,t}$ and uses $z_{t}^{B}$ as a proxy for $\varepsilon_{2,t}$ to remove this correlation. The two assumptions cannot hold at the same time unless $\delta\alpha_{2}^{B}=0$.

Since
\begin{equation*}
\begin{aligned}
	\cov(y_{t+h},z_{t}^{j}) &= \theta_{h,y1}\alpha_{1}^{j} + \theta_{h,y2}\alpha_{2}^{j},\\
	\cov(x_{1,t},z_{t}^{j})&=\alpha_{1}^{j}+\alpha_{2}^{j}\psi_{0,12},\\
	\cov(x_{2,t},z_{t}^{j})&=\alpha_{1}^{j}\psi_{0,21}+\alpha_{2}^{j},
\end{aligned}
\end{equation*}
Assumption~\ref{A-excl}, which sets $\alpha_{2}^{A}=0$, implies
\begin{equation}\label{excl}
	\psi_{0,21}=\frac{\cov(x_{2,t},z_{t}^{A})}{\cov(x_{1,t},z_{t}^{A})},\qquad
	\theta_{h,y1}=\frac{\cov(y_{t+h},z_{t}^{A})}{\cov(x_{1,t},z_{t}^{A})}.
\end{equation}
The second equality says that $\theta_{h,y1}$ is the single-IV estimand $\beta_{h}^{A}$ with $a_{t}^{A}=x_{1,t}$. The covariances with $\tilde{z}_{t}^{A}$ have the same form with $\tilde{\alpha}_{s}^{A}=\alpha_{s}^{A}-\delta\alpha_{s}^{B}$ in place of $\alpha_{s}^{A}$. Assumption~\ref{A-rexcl} sets $\tilde{\alpha}_{2}^{A}=0$ and implies
\begin{equation}\label{rexcl}
	\psi_{0,21}=\frac{\cov(x_{2,t},\tilde{z}_{t}^{A})}{\cov(x_{1,t},\tilde{z}_{t}^{A})},\qquad
	\theta_{h,y1}=\frac{\cov(y_{t+h},\tilde{z}_{t}^{A})}{\cov(x_{1,t},\tilde{z}_{t}^{A})}.
\end{equation}
For $z_{t}^{B}$ as the IV for $\varepsilon_{2,t}$, the two assumptions are the same with $\varepsilon_{2,t}$ as the single shock, $x_{2,t}$ in place of $x_{t}$, and $\tilde{z}_{t}^{B}=z_{t}^{B}-\gamma z_{t}^{A}$ in place of $\tilde{z}_{t}^{A}$, where $\gamma=\cov(z_{t}^{A},z_{t}^{B})/\var(z_{t}^{A})$. By the same steps, they imply
\begin{equation}\label{exclB}
	\psi_{0,12}=\frac{\cov(x_{1,t},z_{t}^{B})}{\cov(x_{2,t},z_{t}^{B})}
	\quad\text{and}\quad
	\psi_{0,12}=\frac{\cov(x_{1,t},\tilde{z}_{t}^{B})}{\cov(x_{2,t},\tilde{z}_{t}^{B})},
\end{equation}
respectively, and identify $\theta_{h,y2}$ as $\cov(y_{t+h},z_{t}^{B})/\cov(x_{2,t},z_{t}^{B})$ and $\cov(y_{t+h},\tilde{z}_{t}^{B})/\cov(x_{2,t},\tilde{z}_{t}^{B})$. 

\section{Application I: The U.S.\ Government Spending Multiplier}\label{Sec-Gov}

The government spending multiplier is the ratio of the change in GDP to the change in government spending. Its magnitude matters for fiscal policy, and whether it exceeds one is still debated. See Blanchard and Perotti (2002), Barro and Redlick (2011), Ramey (2011), Auerbach and Gorodnichenko (2012), Nakamura and Steinsson (2014), and Ramey and Zubairy (2018). Most studies estimate the aggregate or the defense spending multiplier using variation in defense spending. Non-defense spending varies less than defense spending and, more importantly, is likely to be endogenous with respect to GDP (Barro and Redlick, 2011).

We apply our results to the U.S. government spending multiplier in the post-WWII period for two purposes. First, we diagnose the aggregate LP-IV estimands by estimating the weights in Corollary~\ref{C1} for two standard instruments and checking Assumption \textit{Same-Sign}. Second, we illustrate the identification strategy of Section~\ref{Sec-GD} by estimating the defense and non-defense spending multipliers from the two aggregate LP-IV estimates and the sectoral spending series, which serve as the normalizing variables.\footnote{Defense spending is federal defense consumption expenditures and gross investment (FDEFX) from the Federal Reserve Bank of St.\ Louis FRED database, and non-defense spending is total government spending minus defense spending.} The sample period is 1947Q1 to 2015Q4 because quarterly sectoral spending data are available only from 1947.

Following Ramey and Zubairy (2018), we specify the LP-IV regression as 
\begin{equation*}
	\sum_{j=0}^{h}y_{t+j} =\mu_{h} +  \beta_{h}\sum_{j=0}^{h}x_{t+j} + \boldsymbol{\phi}_{h}(L)'\boldsymbol{R}_{t-1}  + u_{t+h},
\end{equation*}
where $y_{t}$ is real GDP, $x_{t}$ is real government spending, $\boldsymbol{R}_{t}$ is a set of control variables, and $\boldsymbol{\phi}_{h}(L)$ is a vector of lag polynomials of order 4. Both $y_{t}$ and $x_{t}$ are divided by trend GDP. The dependent variable and the regressor are sums over horizons $0$ to $h$, so the parameter $\beta_{h}$ is the cumulative government spending multiplier, which is $\widetilde{\beta}_{h}$ of Corollary~\ref{C1}. The instrument is $z_{t}$, and the control variables include lagged values of $y_{t}$, $x_{t}$, and $z_{t}$. The control variables and the construction of trend GDP are also those of Ramey and Zubairy (2018).

We use two IVs in our analysis. The first is the military news shock from Ramey and Zubairy (2018), which we call the RZ news shock. The second is the current defense spending shock from Blanchard and Perotti (2002), which we call the BP defense shock. The BP defense shock is the residual from the regression of current defense spending on the control variables. When using the BP defense shock as an IV, our control variables consist of the lagged values of GDP, total government spending, and defense spending.\footnote{This follows the construction of the Blanchard-Perotti (2002) shock from current government spending in Ramey and Zubairy (2018). The BP defense shock is identified in the same way as the defense spending shock in the SVAR of Blanchard and Perotti (2002, Section IX.B), which includes taxes, defense spending, non-defense spending, and GDP. Our main specification does not include taxes, but the results are similar when taxes are added to the control variables. The BP defense shock may be anticipated because of fiscal foresight (Ramey, 2011), in which case lead-lag exogeneity (Assumption~\ref{A-IV}(iii)) fails. Ramey and Zubairy (2018) note this possibility for the BP total spending shock, which they use as an instrument. Like them, we maintain lead-lag exogeneity.} At the horizons we report, we find no evidence of weak instruments for either IV using the test of Montiel Olea and Pflueger (2013), which accommodates serial correlation in the errors.

\begin{figure}[btp]
	\centering
	\makebox[\textwidth][c]{\includegraphics[width=0.6\linewidth]{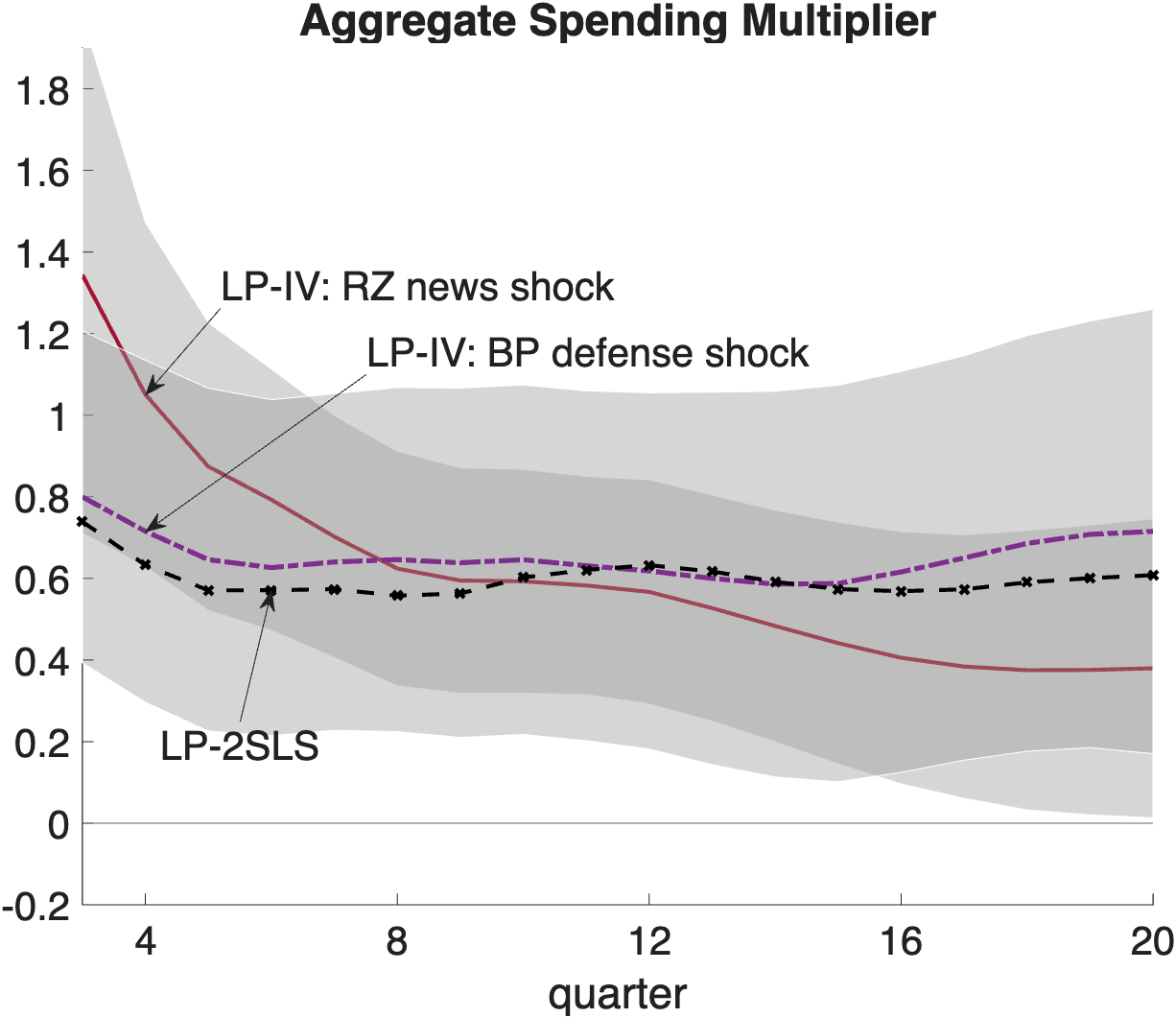}}
	\caption{Cumulative Aggregate Spending Multipliers Across Different Horizons. LP-IV with the RZ news shock (solid), LP-IV with the BP defense shock (dash-dotted), and LP-2SLS with both shocks (dashed with $\times$). The shaded areas are the 95\% confidence bands of the two LP-IV estimates.}
	\label{fig_mul}
\end{figure}

Figure~\ref{fig_mul} presents the cumulative government spending multipliers for horizons from three quarters to five years. Shorter horizons are omitted because the RZ news shock is a weak instrument there. The bands are the pointwise 95\% confidence bands using Newey and West (1987) standard errors with bandwidth equal to the horizon $h$. The LP-IV estimates are based on two different IVs: the RZ news shock and the BP defense shock. These estimates correspond to the left-hand side of \eqref{C1-e1} in Corollary~\ref{C1}. We also report LP-2SLS estimates using both IVs together.

Consistent with Ramey and Zubairy (2018), the estimated multipliers with either the RZ news shock or the BP defense shock are below one after the first year. The estimate with the BP defense shock is smaller than the estimate with the RZ news shock over the first seven quarters and larger afterwards. At five years it is 0.72, against 0.38 with the RZ news shock. The LP-2SLS estimates, which combine both IVs, are close to the LP-IV estimates with the BP defense shock. Each LP-IV estimate can be interpreted as a weighted average of the sectoral multipliers if Assumption~\ref{A-mono} holds for its IV. Even then, the LP-2SLS estimate need not have a structural interpretation, as shown in Section~\ref{Sec-2SLS}. 

We investigate whether Assumption~\ref{A-mono} holds for each IV. Let $s=1$ denote defense spending and $s=2$ denote non-defense spending. By Corollary~\ref{C1}, each aggregate LP-IV estimand is a weighted combination of the cumulative defense and non-defense spending multipliers, and the cumulative version of Assumption \textit{Same-Sign} holds if and only if both weights are nonnegative. Whether the weights can be estimated from the sectoral spending data depends on the relative cumulative inter-component impulse responses of Proposition~\ref{P-aug-cumul}, which we call the (cumulative) inter-sectoral effects. The inter-sectoral effect of a defense spending shock, $\widetilde{\psi}_{h,21}/\widetilde{\psi}_{h,11}$, is the cumulative change in non-defense spending caused by the shock per unit of the cumulative change in defense spending caused by the same shock. The inter-sectoral effect of a non-defense spending shock, $\widetilde{\psi}_{h,12}/\widetilde{\psi}_{h,22}$, is defined in the same way. When both effects are zero, $\widetilde{\theta}_{h,xs}=\widetilde{\psi}_{h,ss}$, so the sectoral multipliers of Corollary~\ref{C1}, $\widetilde{\theta}_{h,ys}/\widetilde{\theta}_{h,xs}$, equal $\widetilde{\theta}_{h,ys}/\widetilde{\psi}_{h,ss}$ and the weights of Corollary~\ref{C1} equal the coefficients $\widetilde{w}_{s}^{j}$ of Proposition~\ref{P-aug-cumul}. They are consistently estimated using sectoral spending data by
\begin{equation*}
	\widehat{w}_{h,s}^{j} = \frac{\sum_{t=1}^{T-h}z_{t}^{\perp(j)}\left(\sum_{\ell=0}^{h}x_{s,t+\ell}^{\perp(j)}\right)}{\sum_{t=1}^{T-h}z_{t}^{\perp(j)}\left(\sum_{\ell=0}^{h}x_{t+\ell}^{\perp(j)}\right)},
\end{equation*}
where $j$ denotes the RZ news shock or the BP defense shock and $v_{t}^{\perp(j)}$ denotes the residual of $v_{t}$ after regressing it on the set of control variables including the lagged values of the instrument $j$.

We examine the decomposition of the cumulative government spending multiplier at $h=18$ when both cumulative inter-sectoral effects are zero. The cumulative component responses $\widetilde{\theta}_{h,ys}/\widetilde{\theta}_{h,xs}$ in the decomposition are estimated by solving the following two equations. The government spending multiplier estimated by the RZ news shock is decomposed as
\begin{equation}
	\begin{array}{ccccccccc}
		\widehat{\beta}^{RZ}_{h} &= & \widehat{w}_{h,1}^{RZ}&  \times& \widehat{\frac{\widetilde{\theta}_{h,y1}}{\widetilde{\theta}_{h,x1}}} &+&  \widehat{w}_{h,2}^{RZ}&\times&\widehat{\frac{\widetilde{\theta}_{h,y2}}{\widetilde{\theta}_{h,x2}}},\\
		0.37& &1.87 &&0.68 & & -0.87 & &1.02 \\
	\end{array}
	\label{dcrz}
\end{equation}
and the multiplier estimated by the BP defense shock is decomposed as
\begin{equation}
	\begin{array}{ccccccccc}
		\widehat{\beta}^{BP}_{h} &= &  \widehat{w}_{h,1}^{BP}&  \times& \widehat{\frac{\widetilde{\theta}_{h,y1}}{\widetilde{\theta}_{h,x1}}} &+&  \widehat{w}_{h,2}^{BP}&\times&\widehat{\frac{\widetilde{\theta}_{h,y2}}{\widetilde{\theta}_{h,x2}}},\\
		0.69 & &0.97 &&0.68 & & 0.03 & &1.02 \\
	\end{array}
	\label{dcbp}
\end{equation}
The decomposition \eqref{dcrz} shows that the multiplier estimated with the RZ news shock is smaller than both component responses because the non-defense weight is negative. Assumption~\ref{A-mono} is therefore violated for the RZ news shock. This raises concerns about the structural interpretation of the multiplier estimated with the RZ news shock. In contrast, both weights in the decomposition \eqref{dcbp} for the BP defense shock are positive, and the non-defense weight is 0.03. The multiplier estimated with the BP defense shock is therefore close to the response to the defense spending shock, 0.69 against 0.68. The signs of the estimated weights are the same at all horizons from three quarters to five years, and the magnitudes are stable after the second year.\footnote{When four lags of defense spending are added to the control variables for the RZ news shock, its non-defense weight remains negative at all of these horizons but is smaller in absolute value, $-0.24$ at $h=18$.}

\begin{figure}[btp]
	\centering
	\makebox[\textwidth][c]{\includegraphics[width=0.9\linewidth]{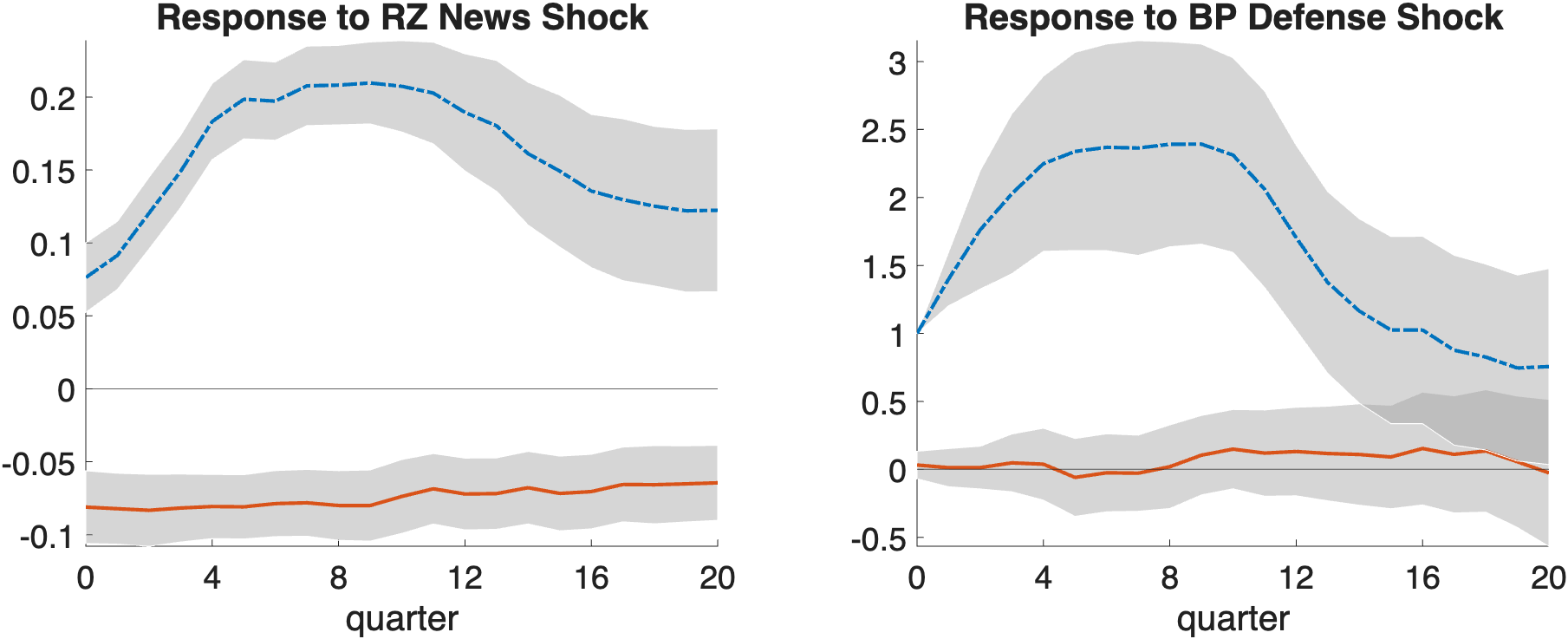}}
	\caption{Impulse Responses of Sectoral Spending to the IV Shocks with 95\% Confidence Bands. Defense spending (dash-dotted) and non-defense spending (solid).}
	\label{fig_irf0}
\end{figure}

Figure~\ref{fig_irf0} shows the responses of defense spending (dash-dotted line) and non-defense spending (solid line) to the RZ news shock (left panel) and to the BP defense shock (right panel). The dependent variables are sectoral spending $x_{s,t+h}$, not cumulative sectoral spending. The response of non-defense spending to the RZ news shock is negative and statistically significant at all horizons, so its cumulative response is negative as well.\footnote{The observation in 1950Q3, when the Korean War began, is influential for the RZ news shock. With a dummy variable for this quarter, the non-defense weight remains negative at all horizons, $-0.31$ at $h=18$, but the response of non-defense spending is no longer statistically significant and the effective F statistic is between 13 and 17.} Under zero inter-sectoral effects, $\cov(\widetilde{x}_{s,t+h},z_{t})=\alpha_{s}\widetilde{\theta}_{h,xs}$, and $\widetilde{\theta}_{h,xs}>0$ because a positive sectoral spending shock raises cumulative total spending. The sign of $\alpha_{s}$ is then the sign of the cumulative response of sector $s$ to the IV. For the RZ news shock, the positive response of defense spending and the negative response of non-defense spending imply $\alpha_{1}>0$ and $\alpha_{2}<0$, so the products $\alpha_{s}\widetilde{\theta}_{h,xs}$ have opposite signs and the cumulative version of Assumption \textit{Same-Sign} is violated. This conclusion does not require the inter-sectoral effects to be exactly zero. Solving $\cov(\widetilde{x}_{r,t+h},z_{t})=\alpha_{1}\widetilde{\psi}_{h,r1}+\alpha_{2}\widetilde{\psi}_{h,r2}$, $r=1,2$, for $\alpha_{2}$ shows that $\alpha_{2}<0$ whenever the inter-sectoral effect of a defense spending shock is above $\widehat{w}_{h,2}^{RZ}/\widehat{w}_{h,1}^{RZ}$, which is $-0.46$ at $h=18$, provided that the product of the two inter-sectoral effects is below one. For the BP defense shock, the response of non-defense spending is small and statistically insignificant at all horizons, which is consistent with $\alpha_{2}$ close to zero and with its near-zero non-defense weight in \eqref{dcbp}.

We now combine the two IVs to estimate the cumulative defense and non-defense spending multipliers, $\widetilde{\theta}_{h,ys}/\widetilde{\psi}_{h,ss}$, using Proposition~\ref{P-aug-cumul}, which requires values of the two cumulative inter-sectoral effects.

For the inter-sectoral effect of a defense spending shock, we use Assumption~\ref{A-excl} of Section~\ref{Sec-GD-excl} with the BP defense shock as $z_{t}^{A}$. The assumption requires the BP defense shock to be uncorrelated with the non-defense spending shock, which holds when defense spending does not respond within the quarter to the non-defense spending shock, as Blanchard and Perotti (2002) assume. The argument of Section~\ref{Sec-GD-excl} applied to cumulative spending then gives $\widetilde{\psi}_{h,21}/\widetilde{\psi}_{h,11}=\cov(\widetilde{x}_{2,t+h},z_{t}^{BP})/\cov(\widetilde{x}_{1,t+h},z_{t}^{BP})$. The estimate of this effect lies between 0.00 and 0.03 across horizons, consistent with the insignificant response of non-defense spending to the BP defense shock in Figure~\ref{fig_irf0}.

No such restriction is available for the inter-sectoral effect of a non-defense spending shock, because neither IV is plausibly uncorrelated with the defense spending shock. We therefore report results for $\widetilde{\psi}_{h,12}/\widetilde{\psi}_{h,22}$ equal to 0, $-0.2$, $-0.4$, and $-0.6$. A value of $-0.2$ means that a non-defense spending shock that raises cumulative non-defense spending by one unit over $h$ quarters lowers cumulative defense spending by 0.2 units over the same quarters. A value of $-1$ is the budget-neutral case, in which every unit of cumulative non-defense spending is offset by a unit of defense spending and total spending is unchanged. The condition $(\widetilde{\psi}_{h,21}/\widetilde{\psi}_{h,11})(\widetilde{\psi}_{h,12}/\widetilde{\psi}_{h,22})\neq1$ of Proposition~\ref{P-aug-cumul} holds for all of these values.

\begin{figure}[btp]
	\centering
	\makebox[\textwidth][c]{\includegraphics[width=0.6\linewidth]{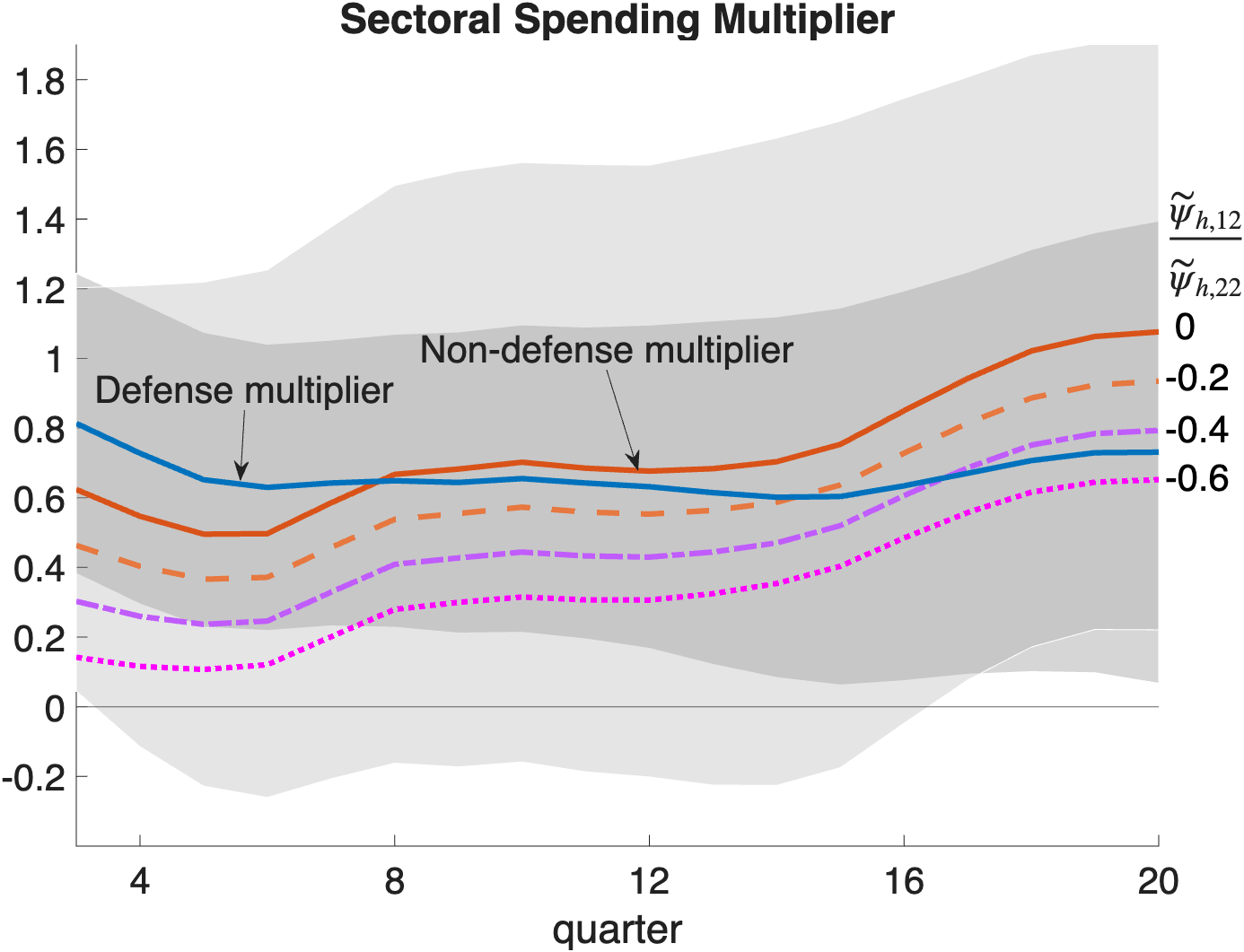}}
	\caption{Cumulative Sectoral Spending Multipliers. The non-defense multiplier is shown for $\widetilde{\psi}_{h,12}/\widetilde{\psi}_{h,22}$ equal to 0, $-0.2$, $-0.4$, and $-0.6$, with the value at the right end of each line, and $\widetilde{\psi}_{h,21}/\widetilde{\psi}_{h,11}$ is fixed at the value implied by Assumption~\ref{A-excl}. The shaded areas are the 95\% confidence bands of the defense multiplier and of the non-defense multiplier at $\widetilde{\psi}_{h,12}/\widetilde{\psi}_{h,22}=0$.}
	\label{fig_sec}
\end{figure}
Figure~\ref{fig_sec} reports the estimated sectoral multipliers.\footnote{$\widetilde{\boldsymbol{b}}_{h}$ in Proposition~\ref{P-aug-cumul} is estimated by the just-identified IV estimator with the two IVs, where the variables in the sample moment condition of each IV have that IV's own control variables partialled out. Its Newey-West covariance matrix gives the confidence band of the non-defense multiplier, which is the second element of $\widetilde{\boldsymbol{b}}_{h}$ at $\widetilde{\psi}_{h,12}/\widetilde{\psi}_{h,22}=0$.} Under Assumption~\ref{A-excl}, the inter-sectoral effect of a defense spending shock is fixed at its estimate (0.00 to 0.03), and the estimated defense multiplier is $\widehat{\beta}_{h}^{BP}/\widehat{w}_{h,1}^{BP}$. This is the single-IV estimate of $\cov(\widetilde{y}_{t+h},z_{t}^{BP})/\cov(\widetilde{x}_{1,t+h},z_{t}^{BP})$, with cumulative defense spending as the regressor and the BP defense shock as the IV. The defense multiplier is stable across horizons, between about 0.6 and 0.8. For the non-defense multiplier, we vary the value of $\widetilde{\psi}_{h,12}/\widetilde{\psi}_{h,22}$. The non-defense multiplier rises from the fourth year for every value of the inter-sectoral effect of a non-defense spending shock, although its confidence bands are wide. At five years it is about 1.1 when the effect is zero, 0.9 at $-0.2$, 0.8 at $-0.4$, and 0.65 at $-0.6$.

\section{Application II: Personal and Corporate Income Tax Multipliers}\label{Sec-MR}

Our second application is the output effect of personal and corporate income tax changes in Mertens and Ravn (2013). They construct narrative measures of legislated changes in the average personal income tax rate (APITR) and the average corporate income tax rate (ACITR), and use them as IVs for the two tax shocks in a proxy SVAR. Following their terminology, we call the two measures the personal proxy and the corporate proxy. A single tax act typically changes both taxes, and the two narrative measures are positively correlated, so each proxy is likely to be correlated with both tax shocks. We apply our results to this setting for two purposes. First, the application shows that the normalizing variables need not be disaggregates of an aggregate. The two tax rates are the normalizing variables. Given the inter-component impact responses, Proposition~\ref{P-aug} identifies the response to each tax shock from the 2SLS regression of the outcome on the two tax rates with the two proxies as the IVs. Each tax shock is normalized to have a unit impact on its own rate. Second, we compute the values of the inter-component impact responses implied by the recursive orderings of Mertens and Ravn (2013) and by the exclusion restrictions, and we show how the estimated response to each tax shock varies with these values and whether these values are economically plausible.

The policy shock is a legislated tax change with a personal component ($s=1$) and a corporate component ($s=2$). The normalizing variables $x_{1,t}$ and $x_{2,t}$ are the average personal and corporate income tax rates. The IVs $z_{t}^{A}$ and $z_{t}^{B}$ are the personal and corporate proxies.

Our specification of the LP-IV regression includes both tax rates as endogenous regressors:
\begin{equation}\label{mrspec}
	y_{t+h} = b_{1,h}x_{1,t} + b_{2,h}x_{2,t} + \boldsymbol{R}_{t}'\boldsymbol{\gamma}_{h} + u_{t+h},
\end{equation}
and the model is estimated by just-identified 2SLS with the two proxies as the IVs. The outcome $y_{t}$ is one of the variables in Figures 2 and 3 of Mertens and Ravn (2013), namely real GDP, each tax rate, each tax base, and government purchases. The tax rates are in percentage points, and GDP, the tax bases, and government purchases are 100 times their logs, so their coefficients are percent responses to a one percentage point rate increase. Revenue responses are constructed from the rate and base responses.\footnote{The revenue response is the rate response divided by the average rate used by Mertens and Ravn (2013), 0.1667 for the personal and 0.2996 for the corporate rate, plus the base response.} The control variables $\boldsymbol{R}_{t}$ consist of a constant and four lags of the two tax rates, the two tax bases, government purchases, GDP, and government debt. Mertens and Ravn (2013) use the same set of variables with the same lags in their VAR.\footnote{Every estimator in this section is a function of covariances between residualized variables, and residualizing the time-$t$ variables on these lags gives the innovations of the VAR system of Mertens and Ravn (2013).} The sample period is 1950Q1 to 2006Q4.

We normalize each tax shock to raise its own tax rate by one percentage point on impact, and we consider several values of the inter-component impact responses. Here $\psi_{0,21}$ is the impact response of the corporate income tax rate to the personal income tax shock, and $\psi_{0,12}$ is the impact response of the personal income tax rate to the corporate income tax shock. We report results for zero inter-component impact responses as a benchmark, for the values implied by the recursive orderings of Mertens and Ravn (2013) in Section~\ref{Sec-GD-ord}, for values in the range $[-0.75,0.75]$, and for the values implied by the exclusion restrictions of Section~\ref{Sec-GD-excl}. Under this benchmark, a tax shock leaves the other average tax rate unchanged within the quarter, while the tax bases and revenues may respond. 

Let $(\theta_{h,y1},\theta_{h,y2})$ be the responses of $y_{t+h}$ to the personal and corporate income tax shocks. By Proposition~\ref{P-aug}, $(\theta_{h,y1},\theta_{h,y2})'=\Psi_{0}'\boldsymbol{b}_{h}$, where $\boldsymbol{b}_{h}=(b_{1,h},b_{2,h})'$ is the 2SLS estimand of \eqref{mrspec}. Row by row,
\begin{equation}\label{crosslines}
	\theta_{h,y1} = b_{1,h} + \psi_{0,21}\,b_{2,h},\qquad
	\theta_{h,y2} = \psi_{0,12}\,b_{1,h} + b_{2,h}.
\end{equation}
The response to the personal income tax shock depends on $\psi_{0,21}$ with slope $b_{2,h}$, and the response to the corporate income tax shock depends on $\psi_{0,12}$ with slope $b_{1,h}$. Under zero inter-component impact responses, the component responses are estimated by $\widehat{\boldsymbol{b}}_{h}$.

\begin{figure}[t]
	\centering
	\includegraphics[width=\textwidth]{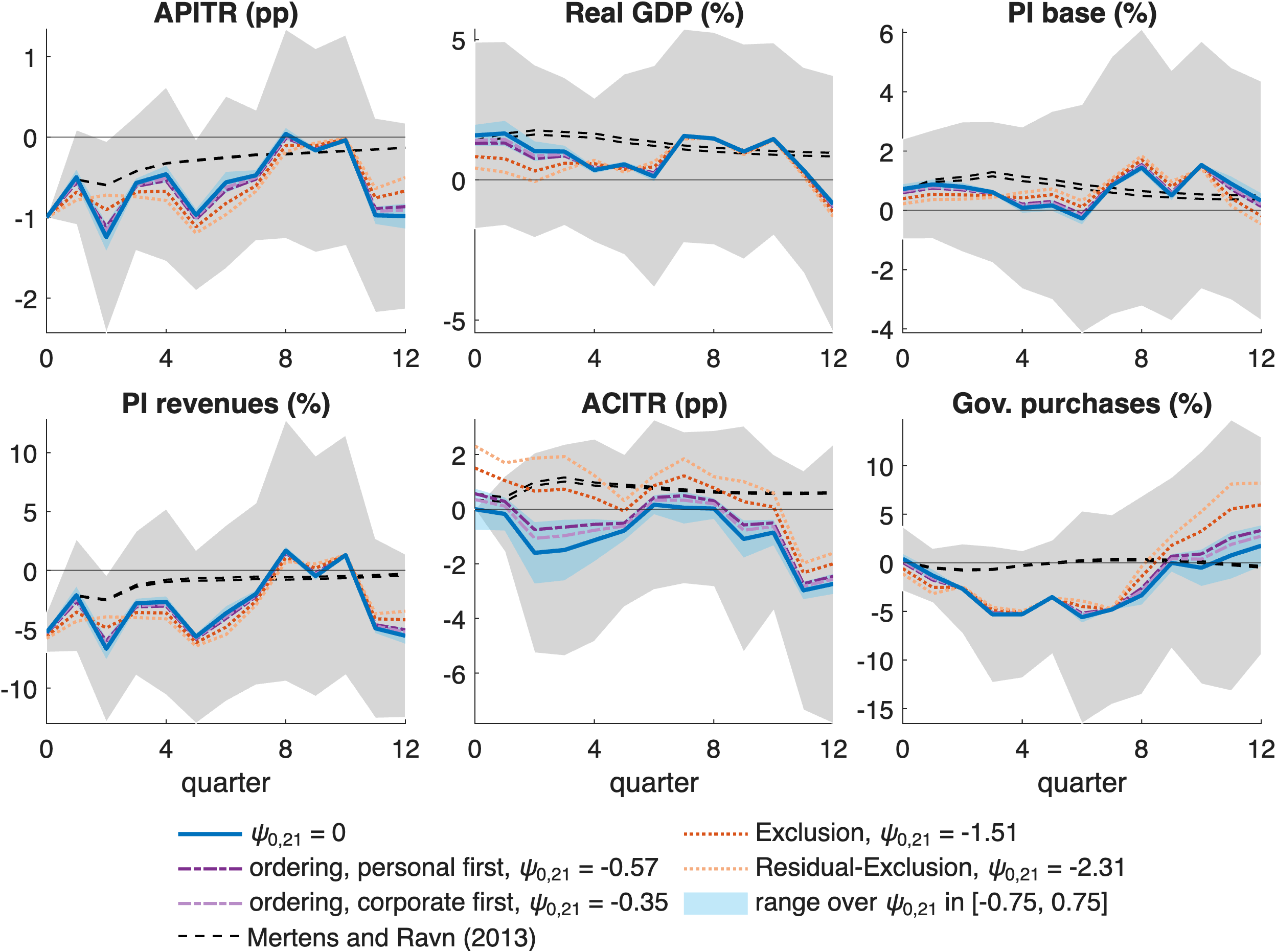}
	\caption{Responses to a One Percentage Point Cut in the Average Personal Income Tax Rate. The gray areas are the 95\% confidence bands of the responses at $\psi_{0,21}=0$, the light blue areas are the range over $\psi_{0,21}$ in $[-0.75,0.75]$, and the legend gives the value of $\psi_{0,21}$ for each line. The variables and units follow Figure 2 of Mertens and Ravn (2013).}
	\label{fig_mr_pi}
\end{figure}

\begin{figure}[t]
	\centering
	\includegraphics[width=\textwidth]{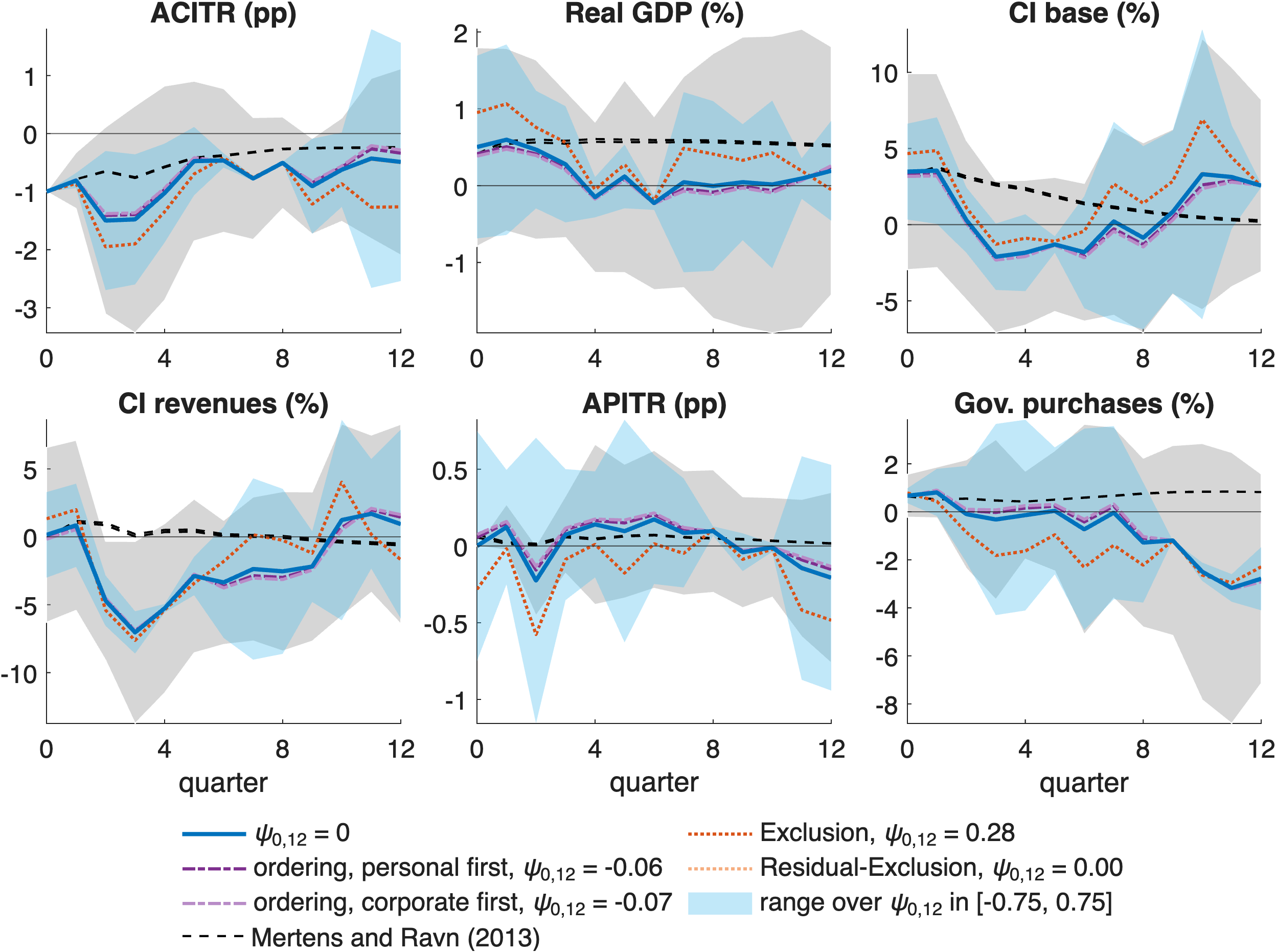}
	\caption{Responses to a One Percentage Point Cut in the Average Corporate Income Tax Rate. The gray areas are the 95\% confidence bands of the responses at $\psi_{0,12}=0$, the light blue areas are the range over $\psi_{0,12}$ in $[-0.75,0.75]$, and the legend gives the value of $\psi_{0,12}$ for each line. The variables and units follow Figure 3 of Mertens and Ravn (2013).}
	\label{fig_mr_ci}
\end{figure}

Figures~\ref{fig_mr_pi} and \ref{fig_mr_ci} report the estimated responses to a personal and a corporate income tax cut for the variables in Figures 2 and 3 of Mertens and Ravn (2013). Each panel shows the responses at $(\psi_{0,12},\psi_{0,21})=(0,0)$ with their 95\% confidence band computed as in Section~\ref{Sec-Gov}, the range of the responses over inter-component impact responses in $[-0.75,0.75]$, and the responses at the values implied by the identifying restrictions. The recursive orderings imply $(\widehat{\psi}_{0,12},\widehat{\psi}_{0,21})=(-0.06,-0.57)$ with the personal income tax shock ordered first and $(-0.07,-0.35)$ with the corporate income tax shock ordered first. Under the two recursive orderings, a personal income tax shock that raises the personal rate by one percentage point lowers the corporate rate by 0.57 or 0.35 percentage points on impact, and a corporate income tax shock that raises the corporate rate by one percentage point lowers the personal rate by 0.06 or 0.07 points. Assumption~\ref{A-excl} imposed on the personal proxy implies $\widehat{\psi}_{0,21}=-1.51$ by \eqref{excl}, and imposed on the corporate proxy it implies $\widehat{\psi}_{0,12}=0.28$ by \eqref{exclB}. This is the case in which Mertens and Ravn (2013, Figure 5) use a single proxy at a time. Assumption~\ref{A-rexcl} imposed on the personal proxy implies $\widehat{\psi}_{0,21}=-2.31$ by \eqref{rexcl}, and imposed on the corporate proxy it implies $\widehat{\psi}_{0,12}=0.00$. All lines are computed from the same $\widehat{\boldsymbol{b}}_{h}$, since replacing a proxy by its projection residual on the other proxy is a nonsingular linear transformation of the two IVs, which leaves the 2SLS estimand unchanged. The two dashed lines are the responses from the proxy SVAR of Mertens and Ravn (2013), one for each ordering, which we replicate. The figures show responses to tax cuts, which are the responses to unit increases multiplied by $-1$. By the unit effect normalization, the own rate falls by one percentage point on impact, and the impact response of the other rate is $-\psi_{0,21}$ in Figure~\ref{fig_mr_pi} and $-\psi_{0,12}$ in Figure~\ref{fig_mr_ci}.

The sensitivity of the responses to the inter-component impact responses follows from \eqref{crosslines}. The response to the personal income tax cut changes with $\psi_{0,21}$ at the rate $-\widehat{b}_{2,h}$, and the response to the corporate income tax cut changes with $\psi_{0,12}$ at the rate $-\widehat{b}_{1,h}$. Table~\ref{tab_mr_b} reports these coefficients for real GDP.\footnote{The responses quoted in the text are computed from the unrounded estimates. The impact coefficients are $-\widehat{b}_{1,0}=1.586$ and $-\widehat{b}_{2,0}=0.503$. To three decimals, the recursive orderings imply $(\widehat{\psi}_{0,12},\widehat{\psi}_{0,21})=(-0.055,-0.567)$ and $(-0.075,-0.355)$, Assumption~\ref{A-excl} implies $\widehat{\psi}_{0,21}=-1.508$ and $\widehat{\psi}_{0,12}=0.282$, and Assumption~\ref{A-rexcl} implies $\widehat{\psi}_{0,21}=-2.314$ and $\widehat{\psi}_{0,12}=0.005$.} The coefficients are imprecisely estimated, so the comparisons below describe the point estimates. Since $\widehat{b}_{2,h}$ is small relative to $\widehat{b}_{1,h}$ at most horizons and near zero beyond a year, the responses in Figure~\ref{fig_mr_pi} are much less sensitive than those in Figure~\ref{fig_mr_ci}. In Figure~\ref{fig_mr_pi}, the impact GDP response is $1.59$ at $\psi_{0,21}=0$, and $1.30$ and $1.41$ at the values implied by the recursive orderings. The values implied by the exclusion restrictions are larger in absolute value, and the impact response falls to $0.83$ at $\psi_{0,21}=-1.51$ (Assumption~\ref{A-excl}) and to $0.42$ at $\psi_{0,21}=-2.31$ (Assumption~\ref{A-rexcl}), while the responses beyond the first year are nearly the same for all values. In Figure~\ref{fig_mr_ci}, the impact GDP response is $0.50$ at $\psi_{0,12}=0$, $0.42$ and $0.38$ at the recursive orderings, $0.95$ at $\psi_{0,12}=0.28$ (Assumption~\ref{A-excl}), and $0.51$ under Assumption~\ref{A-rexcl}, where $\psi_{0,12}$ is close to zero. The differences persist at longer horizons because $\widehat{b}_{1,h}$ remains large at most horizons.

\begin{table}[t]\centering
\caption{Local Projection Coefficients for GDP per One Percentage Point Tax Cut. The entries are $-\widehat{b}_{s,h}$ from \eqref{mrspec}, with Newey-West standard errors in parentheses.}
\label{tab_mr_b}
\begin{tabular}{lccccccc}\toprule
$h$ & 0 & 2 & 4 & 6 & 8 & 10 & 12\\ \midrule
$-\widehat{b}_{1,h}$ & 1.59 & 1.02 & 0.35 & 0.12 & 1.47 & 1.46 & $-0.87$\\
 & (1.70) & (1.57) & (1.31) & (2.02) & (1.93) & (1.75) & (2.34)\\
$-\widehat{b}_{2,h}$ & 0.50 & 0.47 & $-0.15$ & $-0.23$ & 0.00 & 0.02 & 0.19\\
 & (0.66) & (0.60) & (0.50) & (0.58) & (0.88) & (0.99) & (0.82)\\ \bottomrule
\end{tabular}
\end{table}

The values of $\psi_{0,21}$ implied by the exclusion restrictions are too large to be credible. The implied $\psi_{0,21}$ is $-1.51$ under Assumption~\ref{A-excl}, which requires that the personal proxy itself is uncorrelated with the corporate income tax shock, and $-2.31$ under Assumption~\ref{A-rexcl}, which requires that the residual of the personal proxy projected on the corporate proxy is uncorrelated with the corporate income tax shock. Either value means that a personal income tax shock that raises the personal income tax rate by one percentage point lowers the corporate income tax rate by more than one percentage point within the quarter. An inter-component impact response of this size is difficult to justify, which suggests that the exclusion restrictions may be too strong in this application.

The responses reported in Figures~\ref{fig_mr_pi} and \ref{fig_mr_ci} are not as smooth as those reported in Mertens and Ravn (2013), who use a VAR. Local projections estimate each horizon separately, so their responses are typically less smooth. More importantly, the instruments are weak by conventional standards at every horizon. The test of Lewis and Mertens (2026) for 2SLS with multiple endogenous regressors, which is robust to serial correlation, does not reject the null hypothesis of weak instruments at any horizon, whether for both coefficients jointly or for each coefficient separately.\footnote{For the GDP regression, the test statistic is between 0.83 and 0.92 against critical values between 31 and 49, for a bias tolerance of 10\% at the 5\% level under the absolute bias criterion, using their code with the Newey-West option and its default settings otherwise. The regression is just identified, so the test concerns the median bias. The Montiel Olea and Pflueger (2013) effective F statistics of the own-proxy first stages, which test each instrument for its own rate alone, are between 4.2 and 4.9 for the personal proxy and between 1.0 and 1.5 for the corporate proxy, against a threshold of 23.1.} The same first-stage covariances enter the proxy SVAR, so the weakness of the instruments is not specific to local projections. Jentsch and Lunsford (2019) show that the confidence intervals of the proxy SVAR are much wider under a valid bootstrap.

\section{Conclusion}\label{Sec-Conclusion}

This paper characterizes what LP-IV identifies when a valid instrument is correlated with several structural shocks. The estimand is an affine combination of the component impulse responses, with weights that are specific to the instrument and can be negative. The estimand then need not equal any component response or a convex combination of them. For the estimand to be a convex combination of the component responses, the instrument must be correlated with every component shock in the same direction, which is Assumption \textit{Same-Sign}. In the government spending application, under zero inter-sectoral effects, the estimated weight on the non-defense multiplier is negative for the military news instrument of Ramey and Zubairy (2018), which indicates that the instrument is negatively correlated with the non-defense spending shock, so Assumption \textit{Same-Sign} fails. By the equivalence of local projections and SVARs with the instrument ordered first (Plagborg-M\o ller and Wolf, 2021), the same cautions apply to VARs with instruments.

Using several instruments together for the same regressor, as in 2SLS or GMM, does not give the estimand a structural interpretation. The 2SLS estimand is an affine combination of the single-instrument estimands with weights that can themselves be negative, and the over-identifying restrictions test cannot distinguish invalid instruments from a composite policy shock. Identifying the component responses instead requires an observable policy variable for each component shock. With policy variables, as many valid instruments as components, and a restriction on the inter-component impact responses or, for cumulative responses, on their cumulative counterparts, the component responses are identified regardless of the signs of the weights. Section~\ref{Sec-GD} shows that the recursive ordering of a proxy SVAR and exclusion restrictions on the IVs correspond to particular values of the inter-component impact responses, so existing identification strategies can be compared within the same framework.

For applied work, the main implication is that the weights can be estimated from the policy variables under a restriction on the inter-component impact responses, and reporting the weights alongside the LP-IV estimate shows which combination of responses the instrument identifies. The component responses should be reported over a range of values of the inter-component impact responses, as in both applications.

\appendix

	\section{Proofs}
	\label{Sec-pf}
	\subsection{Proof of Proposition~\ref{P1}}
	Write $y_{t+h}$ and $x_{t+h}$ as
	\begin{align*}
		y_{t+h} =& \sum_{j=0}^{\infty}\left(\sum_{s=1}^{S}\theta_{j,ys}\varepsilon_{s,t+h-j} + \sum_{r=S+1}^{m}\theta_{j,yr}\varepsilon_{r,t+h-j} \right),\\
		x_{t+h} =& \sum_{j=0}^{\infty}\left(\sum_{s=1}^{S}\theta_{j,xs}\varepsilon_{s,t+h-j} + \sum_{r=S+1}^{m}\theta_{j,xr}\varepsilon_{r,t+h-j} \right).
	\end{align*} 
	By the bilinearity of covariance and Assumption~\ref{A-IV}, we obtain
	\begin{align*}
		\cov(y_{t+h},z_{t}) =& \sum_{s=1}^{S}\cov(z_{t},\varepsilon_{s,t}) \theta_{h,ys}=  \sum_{s=1}^{S}\alpha_{s}\theta_{h,ys},\\
		\cov(x_{t},z_{t}) =& \sum_{s=1}^{S}\cov(z_{t},\varepsilon_{s,t})\theta_{0,xs} =  \sum_{s=1}^{S}\alpha_{s}\theta_{0,xs}.
	\end{align*}
	Since $\theta_{0,xs}>0$ for $s=1,2,\ldots,S$, the statement of the proposition follows. \qed
	
	\subsection{Proof of Corollary~\ref{C1}}
	The proof is similar to the proof of Proposition~\ref{P1}, and thus omitted.
	
	\subsection{Proof of Proposition~\ref{P-aug}}
	Under the augmented SVMA model \eqref{svma2}, the normalizing variables are written as
\begin{equation*}
	x_{s,t} = \psi_{0,s1}\varepsilon_{1,t} + \psi_{0,s2}\varepsilon_{2,t} + \cdots +\psi_{0,sm}\varepsilon_{m,t} + \{\boldsymbol{\varepsilon}_{t-1},\boldsymbol{\varepsilon}_{t-2},\cdots\},
\end{equation*}
where $\{\cdots\}$ is a linear combination of the elements in the curly brackets. By Assumptions~\ref{A-IV}(ii) and (iii), for $j=A,B$ and $s=1,2$, $\cov(x_{s,t},z_{t}^{j}) = \psi_{0,s1}\alpha_{1}^{j} + \psi_{0,s2}\alpha_{2}^{j}$ where $\alpha_{s}^{j}=\cov(z_{t}^{j},\varepsilon_{s,t})$, and, as in the proof of Proposition~\ref{P1}, $\cov(y_{t+h},z_{t}^{j}) = \alpha_{1}^{j}\theta_{h,y1} + \alpha_{2}^{j}\theta_{h,y2}$. In matrix form, with the unit effect normalization $\psi_{0,11}=\psi_{0,22}=1$,
\begin{equation*}
	C = A\Psi_{0}',\qquad \boldsymbol{c}_{h} = A\boldsymbol{\theta}_{h},\qquad
	A=\left(\begin{array}{cc}\alpha_{1}^{A} & \alpha_{2}^{A}\\ \alpha_{1}^{B} & \alpha_{2}^{B}\end{array}\right),
\end{equation*}
where $\Psi_{0}$ is as defined in Section~\ref{Sec-GD} and $\boldsymbol{\theta}_{h}=(\theta_{h,y1},\theta_{h,y2})'$. Since $\psi_{0,12}\psi_{0,21}\neq1$, $\Psi_{0}'$ is invertible, and since $C$ is invertible, $A=C(\Psi_{0}')^{-1}$ is invertible as well. Therefore
\begin{equation*}
	\boldsymbol{b}_{h}=C^{-1}\boldsymbol{c}_{h}=(\Psi_{0}')^{-1}A^{-1}A\boldsymbol{\theta}_{h}=(\Psi_{0}')^{-1}\boldsymbol{\theta}_{h},
\end{equation*}
and premultiplying by $\Psi_{0}'$ gives the first equality in the proposition. For the second part, let $D=\mathrm{diag}\{\cov(a_{t}^{A},z_{t}^{A}),\cov(a_{t}^{B},z_{t}^{B})\}$, which is nonsingular. Then $W=D^{-1}C$ and $(\beta_{h}^{A},\beta_{h}^{B})'=D^{-1}\boldsymbol{c}_{h}$, so $W^{-1}(\beta_{h}^{A},\beta_{h}^{B})'=C^{-1}DD^{-1}\boldsymbol{c}_{h}=\boldsymbol{b}_{h}$, which gives the second equality. \qed

\section{Comparison with the Potential Outcomes Framework}
\label{Sec-LATE}
				Imbens and Angrist (1994) and Angrist, Imbens, and Rubin (1996) show that the local average treatment effect (LATE) is identified by an instrument under treatment effect heterogeneity. In their potential outcomes framework, the treatment $D_{i}$ is binary and the same within the treated and non-treated groups, while the individual treatment effect $Y_{i}(1)-Y_{i}(0)$ is heterogeneous. 

				In comparison, let us define $\xi_{t}$  as a macroeconomic policy shock which may consist of multiple component shocks $\varepsilon_{s,t}$ in $\boldsymbol{\varepsilon}_{t}$ for $s\in \mathcal{S}_{\xi}\subseteq \{1,2,\ldots,m\}$. Without loss of generality, let $\mathcal{S}_{\xi}=\{1,2,\ldots,S\}$, so that we may write \[\xi_{t}=\sum_{s=1}^{S}\varepsilon_{s,t}.\]
				We refer to $\xi_{t}$ as a composite shock, and it is the unobserved treatment. The ATE corresponds to the impulse response of $y_{t+h}$ to $\xi_{t}$: $\theta_{h,y\xi}\equiv E[y_{t+h}|\xi_{t}=1]-E[y_{t+h}|\xi_{t}=0]$. To illustrate this, assume that the shocks are discrete random variables. Further assuming the convention that $E[y_{t+h}|\xi_{t}=0]$ corresponds to $\boldsymbol{e}_{t}\equiv(\varepsilon_{1,t},\varepsilon_{2,t},\cdots,\varepsilon_{S,t})'=\boldsymbol{0}$, we can write
				\begin{align}
					\nonumber \theta_{h,y\xi} =& \sum_{\boldsymbol{a}\in \{\xi_{t} = 1 \} }\frac{P(\boldsymbol{e}_t=\boldsymbol{a})}{P(\{\xi_{t} = 1 \})}\left(E[y_{t+h}|\boldsymbol{e}_{t}=\boldsymbol{a}]-E[y_{t+h}|\boldsymbol{e}_{t}=\boldsymbol{0}]\right)\\
					\nonumber =& \sum_{\boldsymbol{a}\in \{\xi_{t} = 1 \} }\frac{P(\boldsymbol{e}_t=\boldsymbol{a})}{P(\{\xi_{t} = 1 \})}\boldsymbol{a}'\boldsymbol{\theta}_{h,y}\\
					=& E[\boldsymbol{e}_{t}|\xi_{t}=1]'\boldsymbol{\theta}_{h,y}
					\label{AIR}
				\end{align}
				where $\boldsymbol{a}$ is an $S\times 1$ non-random vector, $\{\xi_{t}=1\}$ is a collection of realizations of $\boldsymbol{e}_{t}$ such that $\xi_{t}=1$, and $\boldsymbol{\theta}_{h,y} = (\theta_{h,y1},\theta_{h,y2},\ldots,\theta_{h,yS})'$. Thus, $\theta_{h,y\xi}$ is a weighted average of component impulse responses and the weights are the conditional means of $\varepsilon_{s,t}$ given $\xi_{t}=1$, which are nonnegative for a large class of distributions of $\boldsymbol{e}_{t}$. For example, if we assume $\varepsilon_{s,t}$ for $s=1,\ldots,S$ are i.i.d., then $E[\varepsilon_{s,t}|\xi_{t}=1]=1/S$ and \eqref{AIR} becomes the equal-weighted average. Unlike in the LATE framework, heterogeneity arises in the composition of $\xi_{t}$, so that the treatment is heterogeneous. Since the component impulse responses are assumed constant, the treatment effect of a particular composition in $\xi_{t}$ is homogeneous.

				The IV provides exogenous variation in the endogenous variable in both cases. However, due to the different nature of the endogeneity (selection versus simultaneity), the way the instrument identifies structural parameters differs between the two frameworks.

				First, consider a binary instrument $Z_{i}$ in the LATE case. The treatment $D_{i}$ is endogenous because it is not randomly assigned. Imbens and Angrist (1994) show that the IV estimand identifies the ATE of the compliers, who receive the treatment if $Z_{i}=1$ but not otherwise. The key identification condition is monotonicity (Condition 2 of Imbens and Angrist, 1994), which rules out defiers, who behave in the opposite way to the compliers. This is a restriction on individual behavior that must be justified in context.

				Now consider our framework. Since $\xi_{t}$ is not observed and its scale is indeterminate, we need to measure the response of $y_{t+h}$ to $\xi_{t}$ relative to the response of another observable endogenous variable $x_{t}$ to $\xi_{t}$. The variable $x_{t}$ is endogenous through simultaneity because the shock $\xi_{t}$ affects $y_{t+h}$ and $x_{t}$ simultaneously. The LP-IV estimand is a convex combination of the component responses only if the correlations between the instrument and the component shocks in $\xi_{t}$ have the same sign. Assumption~\ref{A-mono} plays a role analogous to the monotonicity condition, since it restricts the average relationship between the instrument and the shocks.

				There are a few papers using the potential outcomes framework in the time-series context. Angrist and Kuersteiner (2011) develop semiparametric tests for conditional independence, also known as the unconfoundedness condition. Their framework is different from ours because (i) the treatment is an observed policy action, (ii) the treatment is independent of potential outcomes after conditioning on observables, and as a result, (iii) the role of IV is not discussed. Rambachan and Shephard (2025) develop a general potential outcomes framework for time series that accommodates instruments, and they derive a nonparametric causal interpretation of the LP-IV estimand when the instrument is related to a single treatment. Section~\ref{Sec-ID} discusses how our setting differs, with the instrument correlated with several component shocks.

\end{document}